\documentclass{article}

\usepackage{PRIMEarxiv}

\usepackage[utf8]{inputenc} 
\usepackage[T1]{fontenc}    
\usepackage{hyperref}       
\usepackage{url}            
\usepackage{booktabs}       
\usepackage{amsfonts}       
\usepackage{nicefrac}       
\usepackage{microtype}      
\usepackage{lipsum}
\usepackage{fancyhdr}       
\usepackage{graphicx}       
\graphicspath{{media/}}     
\usepackage{siunitx}
\usepackage{subcaption}  

\usepackage[backend=biber,style=numeric,sorting=none,doi=true,url=false]{biblatex}
\title{Inferring Dislocation Microstructures from X-ray Diffraction via Cross-Modal Contrastive Learning
}

\author{
  Benjamin Udofia\\
  Interdisciplinary Centre for Advanced Materials Simulation\\
  Ruhr University Bochum \\
  Bochum, Germany\\
  \texttt{benjamin.udofia@rub.de} \\
   \And
  Nicolas Bertin \\
  Lawrence Livermore National Laboratory \\
  Livermore \\
  Carlifornia, United States of America\\
  \texttt{bertin1@llnl.gov} \\
  \And
  Markus Stricker\\
  Interdisciplinary Centre for Advanced Materials Simulation\\
  Ruhr University Bochum \\
  Bochum, Germany\\
  \texttt{markus.stricker@rub.de} \\
}

\begin{document}
\maketitle

\begin{abstract}
Understanding and inferring dislocation microstructures from diffraction patterns remains an open challenge in materials characterization, as diffraction measurements provide only indirect information about the underlying dislocation structure.
In this work, a cross-modal learning framework is developed to enable the prediction of 3D dislocation structures directly from diffraction data.
Dislocation density fields generated from discrete dislocation dynamics simulations are paired with corresponding virtual X-ray diffraction patterns and embedded into a shared 2D latent space using contrastive learning. The alignment between structural and diffraction representations of dislocation structures is evaluated directly in the learned latent space using correlations between corresponding latent features.
To estimate the role of dataset size for this approach, farthest point sampling is employed to construct representative and diverse training subsets of varying sizes.
The results show strong cross-modal alignment and that model performance improves rapidly with increasing dataset size.
Near-saturation is achieved with approximately 500 representative observations from a dataset of 10,000 observations, enabling accurate prediction of dislocation density fields from previously unseen diffraction data of the same distribution.
Qualitative comparisons confirm that the predicted structures capture the dominant spatial features of the underlying dislocation microstructures. 
These findings demonstrate an efficient approach for learning structure-diffraction relationships and highlight the potential for inferring structural characteristics of dislocation networks directly from diffraction patterns, providing a pathway toward diffraction-based structural analysis and future extension to experimental data.
\end{abstract}


\section{Introduction}\label{sec1}

Dislocations are the primary carriers of plastic deformation in crystalline materials, and their collective dynamics govern a wide range of macroscopic mechanical properties. Understanding their spatial organization is therefore essential for linking microstructural processes to material behavior. Experimental techniques such as Laue microdiffraction~\cite{ice2009tutorial} provide indirect insights into dislocation evolution through characteristic features such as peak broadening, streaking, and intensity variations. However, directly interpreting diffraction signatures to identify underlying dislocation structures remains an open challenge, with current analyses often reliant on manual interpretation~\cite{kalacska2020investigation}.

Manual analysis of diffraction patterns to identify dislocations is time-consuming and prone to errors~\cite{kalacska2020investigation}.
Borgi et al.~\cite{borgi2025individual} recently demonstrated an approach based on dark-field X-ray microscopy (DFXM) to identify individual dislocations.
While this method is certainly useful it only works for isolated dislocations. 
Such an approach becomes impractical in heavily deformed materials, where complex and dense dislocation networks prove to be a challenge for the unique identification of individual dislocations.

On the computational side, discrete dislocation dynamics (DDD) simulations~\cite{weygand2002aspects, weygand2005study} provide detailed access to dislocation configurations and their 3D evolution in time.
Recent developments have enabled the generation of virtual diffraction patterns from simulated dislocation structures~\cite{bertin2018computation}, establishing a direct link between microstructure and diffraction observables.
In parallel, machine learning approaches have been applied to characterize dislocation microstructures and identify representative structural states~\cite{steinberger2019machine, holm2020overview, udofia2025dislocation}. 
Despite these advances, existing approaches for analyzing diffraction patterns and dislocation microstructures typically treat structural and diffraction data independently and do not explicitly address the inverse mapping from diffraction patterns to dislocation structures.

To address this, we present a cross-modal learning framework to infer dislocation structures directly from diffraction patterns. Dislocation configurations generated from DDD simulations are paired with corresponding virtual X-ray diffraction (XRD) patterns to construct a multimodal dataset. The structural data are represented as dislocation density fields and embedded into a low-dimensional manifold using Isomap~\cite{tenenbaum2000global, udofia2025dislocation}. Farthest point sampling (FPS) is then used to select structurally diverse configurations~\cite{gonzalez1985clustering, moenning2003fast}. 
A contrastive learning model is subsequently trained to learn a shared latent representation between pairs of structural and diffraction modalities~\cite{chen2020simple}. Because the density fields and diffraction patterns provide distinct representations of the same underlying dislocation state, corresponding pairs can be aligned in this shared latent space, enabling the prediction of dislocation structures from diffraction data.

\begin{figure}[htb!]
\centering
\includegraphics[width=\textwidth]{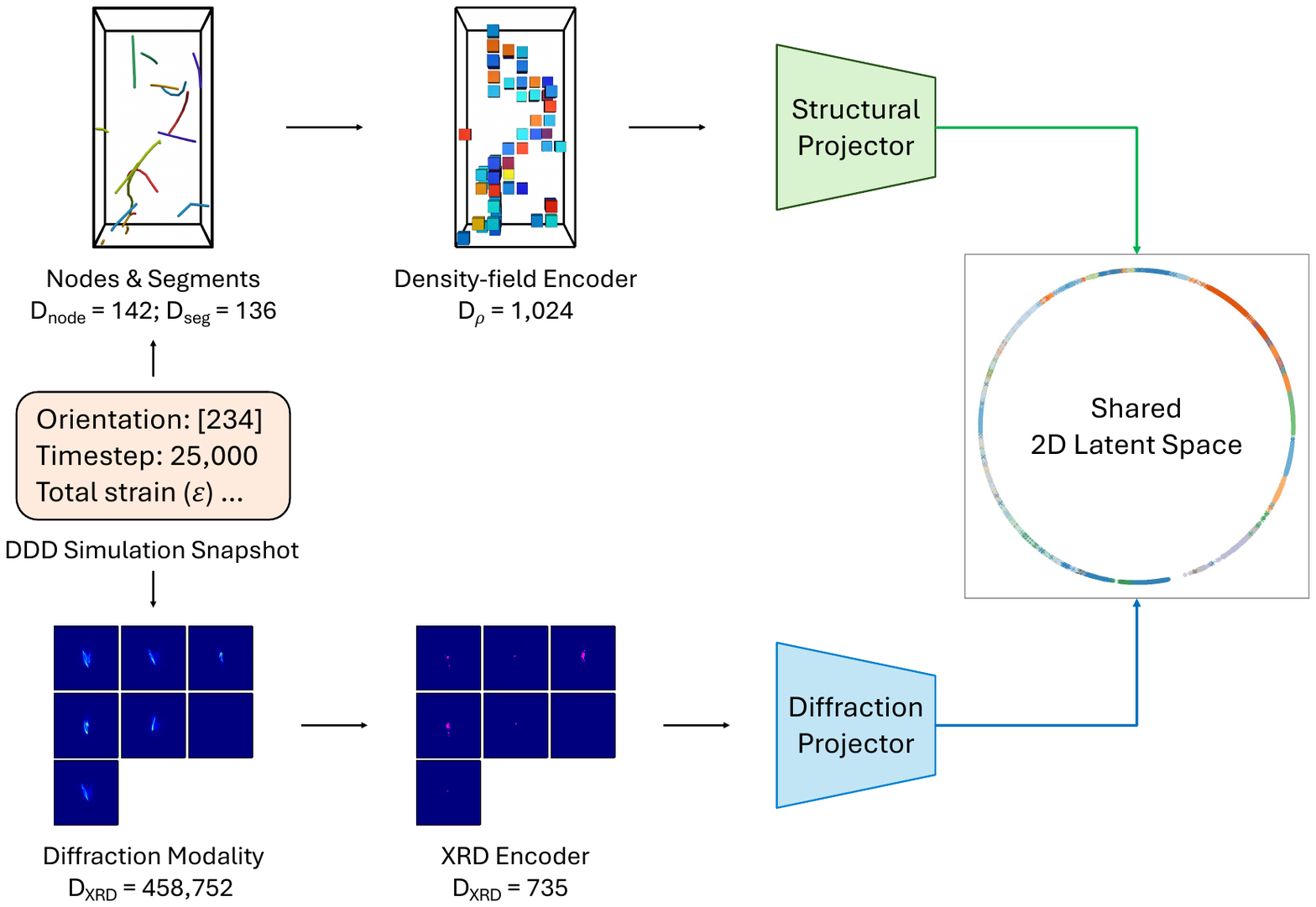}
\caption{
Overview of the simulation-based workflow employed in this study. DDD simulations generate dislocation microstructures represented by nodes and segments. The density-field encoder applies feature engineering by discretizing the dislocation density field to produce a fixed-dimensional structural representation comprising $D_{\rho}=1,024$ features. Corresponding virtual XRD patterns comprising $D_{\mathrm{XRD}}=458,752$ pixels are generated through forward diffraction simulations, after which the XRD encoder applies feature selection based on the variance of the diffraction pixels to retain $735$ informative diffraction features, highlighted in magenta, forming the diffraction representation. Modality-specific structural and diffraction projectors subsequently map both representations into a shared latent space using cross-modal contrastive learning, enabling the learning of structure–diffraction relationships.
}
\label{fig:project_workflow}
\end{figure}

Our framework provides a data-driven approach for addressing the inverse problem of inferring dislocation microstructures from diffraction patterns. The overall simulation-based workflow is illustrated in Figure~\ref{fig:project_workflow}. Dislocation microstructures generated by DDD simulations are initially represented as collections of nodes and segments whose number varies between observations, resulting in structural representations of variable dimensionality that are not directly suitable for machine learning. The density-field encoder therefore discretizes the dislocation density field on a regular voxel grid, transforming these variable-length dislocation configurations into a fixed-dimensional structural representation suitable for machine learning. In parallel, corresponding virtual XRD patterns are generated through forward diffraction simulations~\cite{bertin2018computation}. Because the diffraction patterns are inherently high-dimensional, the XRD encoder applies feature selection \cite{guyon2003introduction}, using the variance of the diffraction pixels to identify the most informative features and form a compact diffraction representation. The resulting structural and diffraction representations are subsequently mapped by modality-specific projection heads into a shared 2D latent space using cross-modal contrastive learning~\cite{chen2020simple}, where corresponding dislocation microstructures and diffraction patterns are aligned. In the present study, all training, testing, and validation data are generated from simulations, enabling evaluation of the proposed framework under well-defined conditions. By demonstrating that dislocation microstructures can be inferred from diffraction signatures using a limited number of representative samples, this work establishes a foundation for future integration with experimental diffraction measurements and automated diffraction-based microstructure characterization.

\section{Methods}
This section describes the computational workflow used to construct the multimodal dataset linking dislocation microstructures and their diffraction signatures. Dislocation configurations are generated using DDD simulations and represented as density fields. From these configurations, corresponding virtual XRD patterns are computed. These two modalities form the basis for the cross-modal learning framework developed in this work.

\subsection{Discrete Dislocation Dynamics Simulations} \label{sec:DDD}
Dislocation microstructures are generated using discrete dislocation dynamics (DDD) simulations performed with the Karlsruhe DDD code under open boundary conditions~\cite{weygand2002aspects, weygand2005study}, consistent with small-scale plasticity experiments on micropillars~\cite{greer2006nanoscale, tang2007dislocation}.
DDD provides direct access to 3D dislocation configurations that are difficult to obtain experimentally in heavily deformed materials and therefore serves as the basis for establishing structure-diffraction relationships.

Following the procedure established in our previous work~\cite{udofia2025dislocation}, simulations are performed for an fcc crystal with material properties representative of aluminum (lattice constant $a=\SI{0.404}{\nano\meter}$, Poisson's ratio $\nu=0.347$, and shear modulus $G=\SI{27}{\giga\pascal}$). The simulation volume measures $240 \times 480 \times 240$~\si{\nano\metre\cubed} and is initialized with randomly distributed Frank-Read sources of nominal length \SI{100}{\nano\meter} and an initial dislocation density of \SI{2.5e13}{\per\meter\squared}.

Uniaxial compression is applied using displacement control at a constant strain rate of \SI{5000}{\per\second}, consistent with typical DDD simulations and motivated by the widespread use of compression testing in small-scale plasticity experiments~\cite{frick2010plasticity, maass2012situ, weekes2015situ, malyar2017size}. Compression is imposed along the $[100]$, $[110]$, $[111]$, and $[234]$ crystallographic directions. To account for experimentally relevant deviations from ideal loading conditions, additional simulations are performed with loading-axis misorientations of $\pm$\SI{5}{\degree} and $\pm$\SI{10}{\degree} relative to each nominal orientation. These loading conditions activate different combinations of slip systems and generate a broader range of dislocation microstructures, mimicking experimental variability w.r.t. the alignment between the crystallographic axis and nominal compression direction. Simulations are continued to a total compressive strain of approximately 1\%, and dislocation configurations are recorded throughout the deformation process. Each retained configuration is treated as an individual \textit{observation} of a \textit{dislocation structure state}, independent of its position within a particular simulation trajectory. For each loading condition, 500 configurations are retained, resulting in a dataset of 10,000 observations~\cite{udofia2024-11354118}.

The discrete dislocation networks are converted into voxelized dislocation density fields using a discretization of $8 \times 16 \times 8$. This representation transforms the discrete line-based dislocation structure into a fixed-dimensional field description containing 1,024 features, while preserving the spatial distribution of dislocation content within the simulation volume~\cite{udofia2025dislocation}. These density fields constitute the structural representation used throughout the subsequent manifold learning and contrastive learning analyses.

In the present work, two complementary modalities are considered: a structural modality represented by discretized dislocation density fields and a diffraction modality represented by virtual X-ray diffraction patterns. Both modalities originate from the same underlying dislocation microstructure and are paired for cross-modal learning.

\subsection{Virtual X-ray Diffraction Generation}
The diffraction modality is represented by virtual X-ray diffraction (XRD) patterns generated from the dislocation configurations obtained from the DDD simulations. 
Virtual XRD patterns are calculated using the DDD-based X-ray diffraction (DDD-XRD) framework developed by Bertin et al.~\cite{bertin2018computation}. For each dislocation configuration, the deformation-gradient field induced by the discrete dislocation network is evaluated on a regular sampling grid. Its symmetric and antisymmetric components describe the local elastic strain and lattice rotation, respectively, which modify the local diffraction condition. The corresponding diffracted rays are subsequently ray-traced for each sampling point, and their intensity contributions are integrated on a virtual detector to generate the final diffraction pattern.

Because the DDD-XRD framework takes dislocation network data in the ParaDiS~\cite{arsenlis2007enabling} format as input, the output of the Karlsruhe DDD code is first converted into the ParaDiS dislocation network format. The conversion preserves dislocation-node positions, segment connectivity, Burgers vectors, and slip-plane normals required for the diffraction calculation. 
This conversion enables virtual diffraction calculations for dislocation configurations generated with the Karlsruhe DDD code under the open-boundary conditions employed in the present simulations. Free-surface boundary conditions are specified in the DDD-XRD calculation, consistent with the open-boundary conditions used in the DDD simulations.

To account for the non-cubic simulation volume of $240 \times 480 \times 240$~\si{\nano\metre\cubed}, the DDD-XRD implementation is adapted to use a rectangular sampling grid of $64 \times 128 \times 64$ points. Virtual Laue diffraction patterns are calculated using a fixed incident beam direction ($s_0=[001]$), detector orientation ($D=[100]$), and detector distance of \SI{70}{\milli\metre}. The incident X-ray spectrum spans energies from \SI{5}{\kilo\electronvolt} to \SI{25}{\kilo\electronvolt}. Diffraction intensities are recorded on virtual detectors with a resolution of $256 \times 256$ pixels. Seven reflections are considered: [111], [113], [13$\bar{1}$], [131], [204], [220], and [224]. These reflections are selected to capture diffraction features associated with the diverse dislocation microstructures generated by the different loading orientations and activated slip systems. Together, they provide a representative sampling of the diffraction signatures present in the dataset.

The resulting diffraction representation contains 458,752 intensity features per observation. To reduce its dimensionality while retaining the most informative diffraction features, feature selection~\cite{guyon2003introduction} is performed using the variance of the diffraction pixels across the dataset as the selection criterion. The feature-selection process and the resulting diffraction representation are presented in Section~\ref{sec:results}.

\section{Machine Learning Methodologies}\label{sec2}
This section outlines the machine learning techniques employed to analyze and model the relationship between dislocation microstructures and their corresponding diffraction patterns. The methodology integrates manifold learning for low-dimensional representation~\cite{udofia2025dislocation}, FPS for selecting structurally diverse subsets~\cite{gonzalez1985clustering, moenning2003fast}, and cross-modal contrastive learning to learn a shared latent space between structural and diffraction data~\cite{chen2020simple,radford2021learning}.

\subsection{Manifold Learning}
The voxelized dislocation density fields contain 1,024 features per configuration, making direct comparison and systematic sampling of structural similarity difficult. Manifold learning is therefore used to represent these high-dimensional microstructures in a lower-dimensional embedding that preserves their relative structural relationships~\cite{udofia2025dislocation}. 

In this work, Isomap, as implemented in the \texttt{scikit-learn} library~\cite{scikit-learn}, is employed to obtain a low-dimensional embedding of the dislocation density fields. Isomap extends classical multidimensional scaling (MDS)~\cite{kruskal1964nonmetric, borg2005modern} by preserving geodesic distances along the data manifold rather than Euclidean distances in the original high-dimensional space~\cite{tenenbaum2000global}. A neighborhood graph is first constructed by connecting each data point to its nearest neighbors, with edges weighted by pairwise distances. The geodesic distance between two points is then approximated as the shortest path distance on this graph.

Let $\mathbf{x}_i \in \mathbb{R}^D$, $i = 1, \dots, N$, denote the high-dimensional data points, and let $\mathbf{y}_i \in \mathbb{R}^d$ denote their corresponding low-dimensional embeddings, with $d \ll D$. The pairwise geodesic distance matrix $D_G(i,j)$ is computed using shortest path algorithms on the neighborhood graph. The low-dimensional embedding $\mathbf{y}_i$ is then obtained by minimizing the MDS objective~\cite{borg2005modern}
\begin{equation}
\sum_{i<j} \left( D_G(i,j) - \| \mathbf{y}_i - \mathbf{y}_j \| \right)^2, \label{eq1}
\end{equation}
which preserves the manifold geometry in the reduced space. The resulting embedding provides a compact representation of the structural relationships within the simulated dataset. Distances in this embedding space are subsequently used by FPS to identify structurally diverse configurations for the cross-modal learning dataset.

\subsection{Farthest Point Sampling}
The simulated dataset contains dislocation configurations that may be structurally similar despite originating from different loading conditions or deformation stages. In particular, consecutively recorded configurations can provide redundant observations of nearby dislocation states. Because it is neither computationally feasible nor necessary to include every possible dislocation configuration in the training set, FPS~\cite{gonzalez1985clustering, moenning2003fast}
is used to select a representative subset to provide broad coverage of the structural variability contained in the simulated dataset.

FPS is applied to the low-dimensional Isomap embedding of the dislocation density fields. It is a greedy sampling algorithm that iteratively selects the configuration whose minimum distance to the previously selected set is maximal~\cite{gonzalez1985clustering, moenning2003fast}. Consequently, each newly selected configuration is structurally dissimilar from the samples already retained, promoting coverage of the structural manifold while reducing redundancy. The corresponding virtual XRD patterns are selected together with their paired density fields and are subsequently used to train the cross-modal learning model.

Given a dataset $\mathcal{X} = \{\mathbf{y}_i\}_{i=1}^N$, where $\mathbf{y}_i$ denotes the low-dimensional embedding of the $i$th dislocation configuration obtained from manifold learning, the sampling procedure begins by randomly selecting one point to form the initial set $S_1$. At each iteration $k \geq 2$, the next selected point $\mathbf{y}_k$ is chosen according to
\begin{equation}
\mathbf{y}_k = \arg\max_{\mathbf{y} \in \mathcal{X} \setminus S_{k-1}} \; \min_{\mathbf{y}_j \in S_{k-1}} \| \mathbf{y} - \mathbf{y}_j \|,
\label{eq2}
\end{equation}
where $S_{k-1}$ denotes the set of the $k-1$ previously selected points, $\mathbf{y}_j$ denotes a previously selected embedding in $S_{k-1}$, and $\|\cdot\|$ denotes the Euclidean norm in the embedding space. This procedure promotes the selection of points that are maximally separated from each other, enabling efficient coverage of the structural manifold. In this work, FPS is applied to the Isomap embeddings of the dislocation density fields to construct subsets of varying sizes for subsequent analysis.

\subsection{Cross-Modal Contrastive Learning}
Cross-modal contrastive learning learns a shared latent space between different but related data modalities by bringing representations of corresponding data pairs closer together while pushing representations of unrelated pairs further apart within the latent space~\cite{chen2020simple, radford2021learning}. In the present work, the structural and diffraction modalities provide different representations of the same underlying dislocation microstructure: the density field describes the spatial distribution of dislocation content, whereas the diffraction pattern reflects the associated lattice distortion integrated over the volume. Because these modalities differ substantially in their physical meaning and dimensionality, direct comparison between them is not straightforward. The paired simulation data therefore provide positive structure-diffraction examples of the same dislocation state, whereas density fields and diffraction patterns originating from different observations provide negative examples.

Rather than explicitly inverting the forward diffraction calculation performed by the DDD-XRD framework~\cite{bertin2018computation}, the model learns the structure-diffraction correspondence represented by the simulated dataset. This approach is motivated by the fact that the inverse mapping is non-unique, as distinct dislocation configurations can produce similar diffraction signatures. The objective is therefore to learn a shared latent space in which corresponding structure-diffraction pairs (positive pairs) are mapped close together, while unrelated pairs (negative pairs) remain well separated.

Let $\mathbf{x}_i$ and $\mathbf{z}_i$ denote the dislocation density field and the corresponding diffraction pattern for the $i$-th observation, respectively. The cross-modal contrastive learning model employs two-layer multilayer perceptrons (MLPs) to map the structural and diffraction representations into L2-normalized latent representations $\mathbf{h}_i^{(x)}$ and $\mathbf{h}_i^{(z)}$. Each MLP consists of a fully connected layer with 256 hidden units, followed by a ReLU activation and a fully connected output layer producing a latent representation of dimension $d=2$. This latent dimension is chosen to enable direct visualization of the cross-modal alignment while maintaining a low-dimensional representation. The correspondence between the structural and diffraction representations is quantified using cosine similarity,

\begin{equation}
\text{sim}(\mathbf{h}_i^{(x)}, \mathbf{h}_j^{(z)}) =
\frac{\mathbf{h}_i^{(x)} \cdot \mathbf{h}_j^{(z)}}
{\|\mathbf{h}_i^{(x)}\| \, \|\mathbf{h}_j^{(z)}\|} 
=
\mathbf{h}_i^{(x)} \cdot \mathbf{h}_j^{(z)}, \label{eq3}
\end{equation}
where $i,j = 1, \dots, B$ index samples within a batch; the denominator equals one because the latent representations are L2-normalized.

The model is trained using a contrastive loss based on the InfoNCE objective, implemented as a temperature-scaled cross-entropy loss over pairwise similarities within each batch~\cite{oord2018representation, chen2020simple},
\begin{equation}
\mathcal{L} = - \frac{1}{B} \sum_{i=1}^{B}
\log \frac{\exp\left( \text{sim}(\mathbf{h}_i^{(x)}, \mathbf{h}_i^{(z)}) / \tau \right)}
{\sum_{j=1}^{B} \exp\left( \text{sim}(\mathbf{h}_i^{(x)}, \mathbf{h}_j^{(z)}) / \tau \right)}. \label{eq4}
\end{equation}
Here, $B$ denotes the batch size and $\tau$ is a temperature parameter that controls the concentration of the similarity distribution. This loss encourages alignment of positive pairs while maintaining separation between negative pairs.
The learned shared latent space enables cross-modal retrieval and provides the basis for predicting structural representations from diffraction patterns.

\section{Results}\label{sec:results}
In this section, we evaluate the ability of our contrastive learning framework to associate dislocation microstructures with their corresponding simulated XRD signatures. We begin by characterizing the DDD-generated dislocation configurations and analyzing their low-dimensional embeddings to quantify structural similarity between dislocation configurations. FPS is then applied to these embeddings to obtain structurally diverse yet representative subsets of dislocation configurations. Building on these structural representations, we analyze the corresponding virtual diffraction dataset and perform feature selection \cite{guyon2003introduction} to identify the diffraction pixels exhibiting the greatest variability across the simulated dataset. We then assess the performance of the cross-modal contrastive model in aligning structural and diffraction representations in the shared latent space and quantify predictive accuracy using similarity-based metrics. Finally, we examine the robustness and generalization capabilities of the model through cross-validation strategies.

\subsection{Dislocation Microstructure Dataset}
\label{sec:results_dataset}
The 10,000 dislocation density-field observations derived from DDD simulations are embedded into a low-dimensional structural manifold using the Isomap algorithm~\cite{scikit-learn}. The Isomap embedding is generated using the same hyperparameters reported in Udofia et al.~\cite{udofia2025dislocation}.

\begin{figure}[htbp]
    \centering
    \begin{subfigure}[t]{0.25\textwidth}
        \centering
        \includegraphics[width=\textwidth]{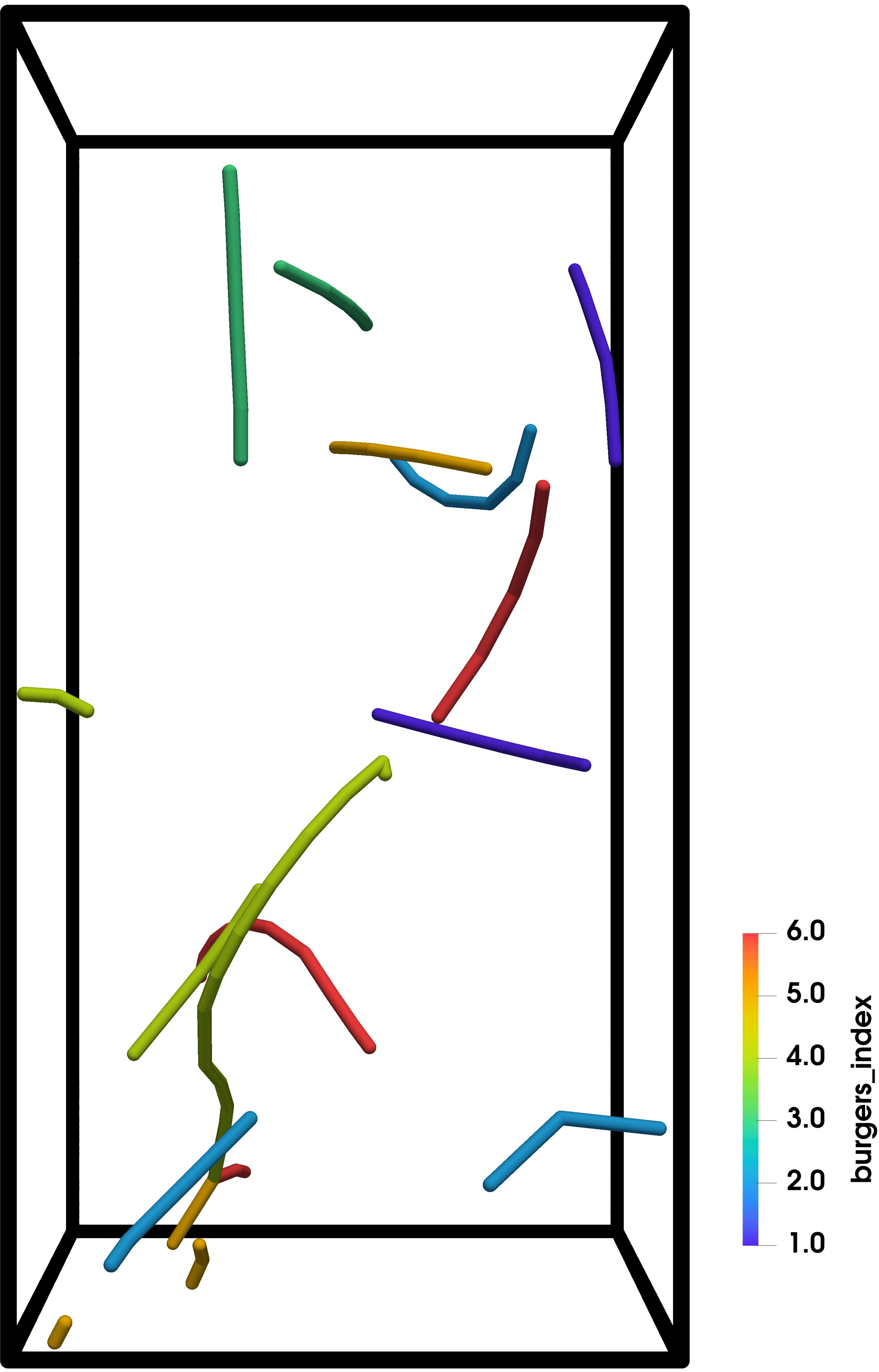}
        \caption{Representative dislocation configuration under compression.}
        \label{fig:ddd_config}
    \end{subfigure}
    \hfill
    \begin{subfigure}[t]{0.27\textwidth}
        \centering
        \includegraphics[width=\textwidth]{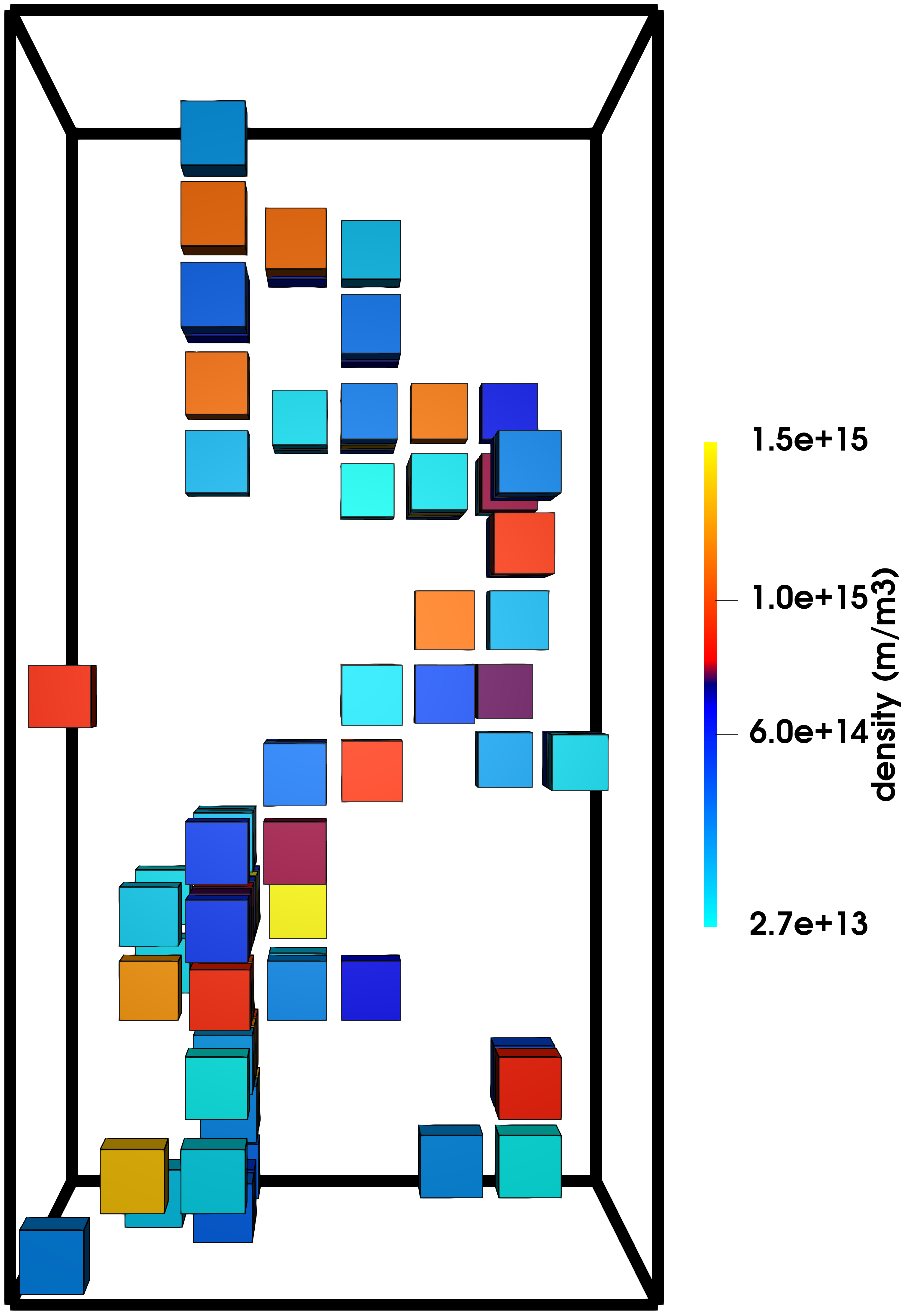}
        \caption{Corresponding discretized dislocation density field.}
        \label{fig:density_field}
    \end{subfigure}
    \hfill
    \begin{subfigure}[t]{0.4\textwidth}
        \centering
        \includegraphics[width=\textwidth]{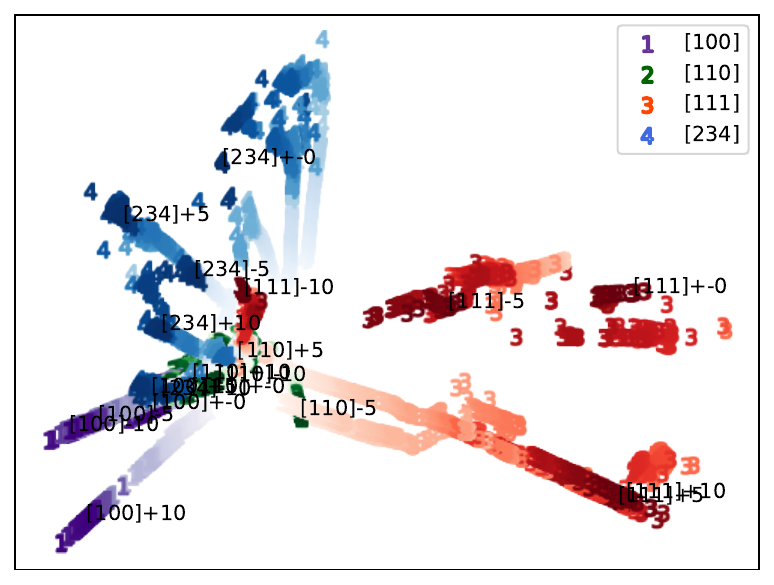}
        \caption{Isomap embedding of the nominal loading orientations and corresponding $\pm\SI{5}{\degree}$ and $\pm\SI{10}{\degree}$ misorientations.}
        \label{fig:isomap_embedding}
    \end{subfigure}
    \caption{
    Workflow for generating low-dimensional structural representations from DDD simulations. The process proceeds sequentially from (a) dislocation line networks, to (b) discretized dislocation density fields, and finally to (c) the Isomap embedding of the complete dataset.
    }
    \label{fig:embedding_workflow}
\end{figure}

Figure~\ref{fig:embedding_workflow} illustrates the transformation from DDD-generated dislocation microstructures to their corresponding low-dimensional representation. A representative DDD configuration (Figure~\ref{fig:ddd_config}) shows dislocation line segments colored according to their Burgers vector, highlighting the crystallographic character of the dislocation lines. The corresponding discretized density field (Figure~\ref{fig:density_field}) provides a structured representation of the dislocation network on a regular $8 \times 16 \times 8$ voxel grid, enabling quantitative comparison across observations. The resulting Isomap embedding (Figure~\ref{fig:isomap_embedding}) organizes the microstructures according to structural similarity. 

Distinct clusters corresponding to the four nominal loading orientations ($[100]$, $[110]$, $[111]$, and $[234]$) remain visible in the low-dimensional space. The additional loading-axis misorientations ($\pm \SI{5}{\degree}$ and $\pm \SI{10}{\degree}$) are represented within these clusters, consistent with the structural relationships reported previously~\cite{udofia2025dislocation}. Within each orientation group, the embedded configurations form a continuous trajectory that reflects the evolution of the dislocation structures with increasing strain. The color gradient in the embedding visualization in Figure~\ref{fig:isomap_embedding} further highlights this trajectory, indicating that the manifold captures both orientation-dependent and strain-dependent variations in the microstructure.

This low-dimensional manifold therefore provides a compact and physically meaningful representation of the dislocation dataset. In the present work, individual embedded configurations are treated as observations of dislocation states rather than as trajectories of individual simulations, providing the structural basis for subsequent farthest point sampling and cross-modal learning analyses.

\subsection{Representative Subset Selection}
The Isomap embedding represents each dislocation microstructure as a point on a low-dimensional manifold, with structurally similar observations appearing close to one another. 
To reduce redundancy and therefore oversampling while preserving structural diversity, FPS is applied to the embedding coordinates. At each iteration, the algorithm selects the observation whose embedding is farthest from those of the previously selected observations, resulting in a subset that is well distributed across the embedding and provides broad coverage of the structural manifold.

Sampling is performed with increasing numbers of observations, with representative sets containing 10, 50, 100, 500, 1000, 2000, 3000, 5000, 7000, and 9000 observations. Smaller subsets provide a coarse representation of the manifold, capturing its global organization, while larger subsets progressively recover finer structural variations.

\begin{figure}[htbp]
    \centering
    \begin{subfigure}[b]{0.32\textwidth}
        \centering
        \includegraphics[width=\textwidth]{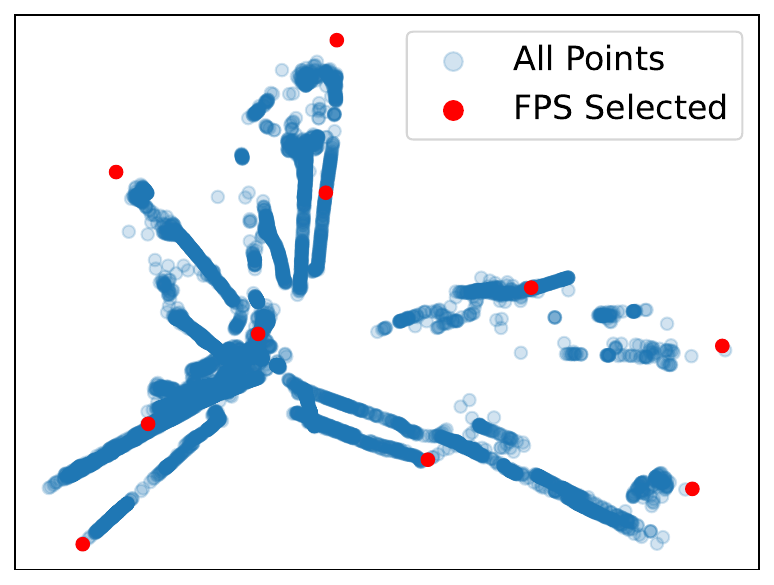}
        \caption{Coarse sampling: 10 observations}
        \label{fig:fps_10}
    \end{subfigure}
    \hfill
    \begin{subfigure}[b]{0.32\textwidth}
        \centering
        \includegraphics[width=\textwidth]{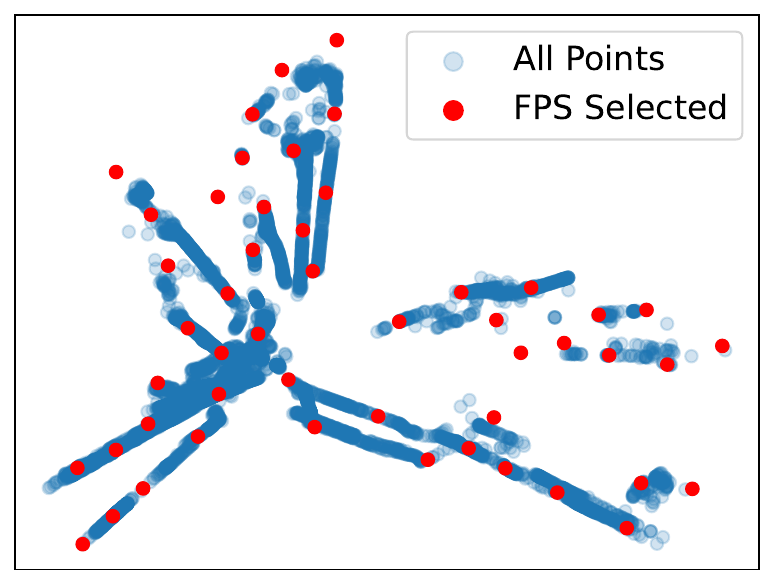}
        \caption{Intermediate sampling: 50 observations}
        \label{fig:fps_50}
    \end{subfigure}
    \hfill
    \begin{subfigure}[b]{0.32\textwidth}
        \centering
        \includegraphics[width=\textwidth]{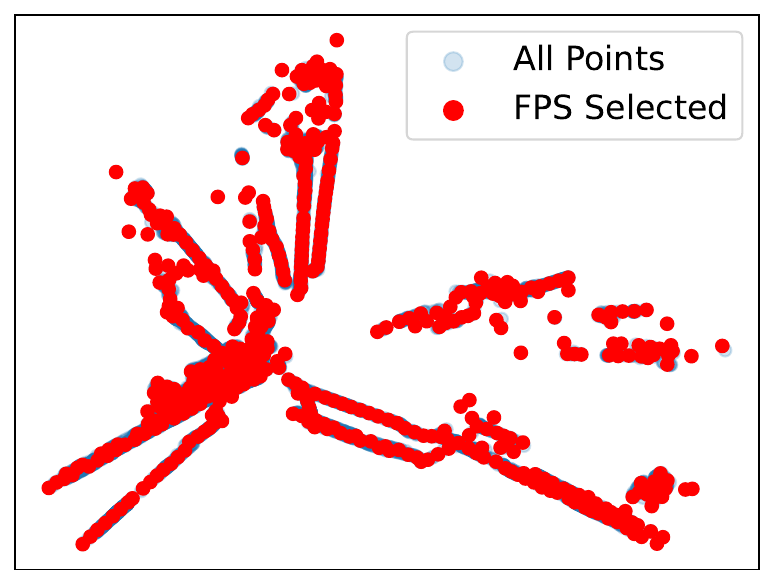}
        \caption{Fine sampling: 500 observations}
        \label{fig:fps_300}
    \end{subfigure}
    \caption{
    Farthest point sampling (FPS) applied to the Isomap embedding of the dislocation microstructure dataset.
    (a) Coarse sampling with 10 selected data points capturing the broad structure of the manifold. 
    (b) Intermediate sampling with 50 data points providing improved coverage of the embedding space. 
    (c) Fine sampling with 500 data points closely approximating the full dataset distribution.
    }
    \label{fig:fps}
\end{figure}

Figure~\ref{fig:fps} illustrates the effect of FPS on the Isomap embedding. Coarse sampling (Figure~\ref{fig:fps}a) selects 10 observations that span the overall structure of the manifold but leave large regions unsampled. Intermediate sampling (Figure~\ref{fig:fps}b) selects 50 observations, improving coverage by sampling additional regions of the embedding space. Fine sampling (Figure~\ref{fig:fps}c) selects 500 observations, producing a dense representation that closely approximates the distribution of the full dataset while resolving finer structural variations.

The FPS-selected observations constitute the representative structural subset used to construct paired diffraction data and train the contrastive learning model.

\subsection{Multimodal Data Integration}
\label{sec:multimodal_integration}
\begin{figure}[htb!]
\centering
\begin{subfigure}[t]{0.45\textwidth}
    \centering
    \includegraphics[height=5.3cm]{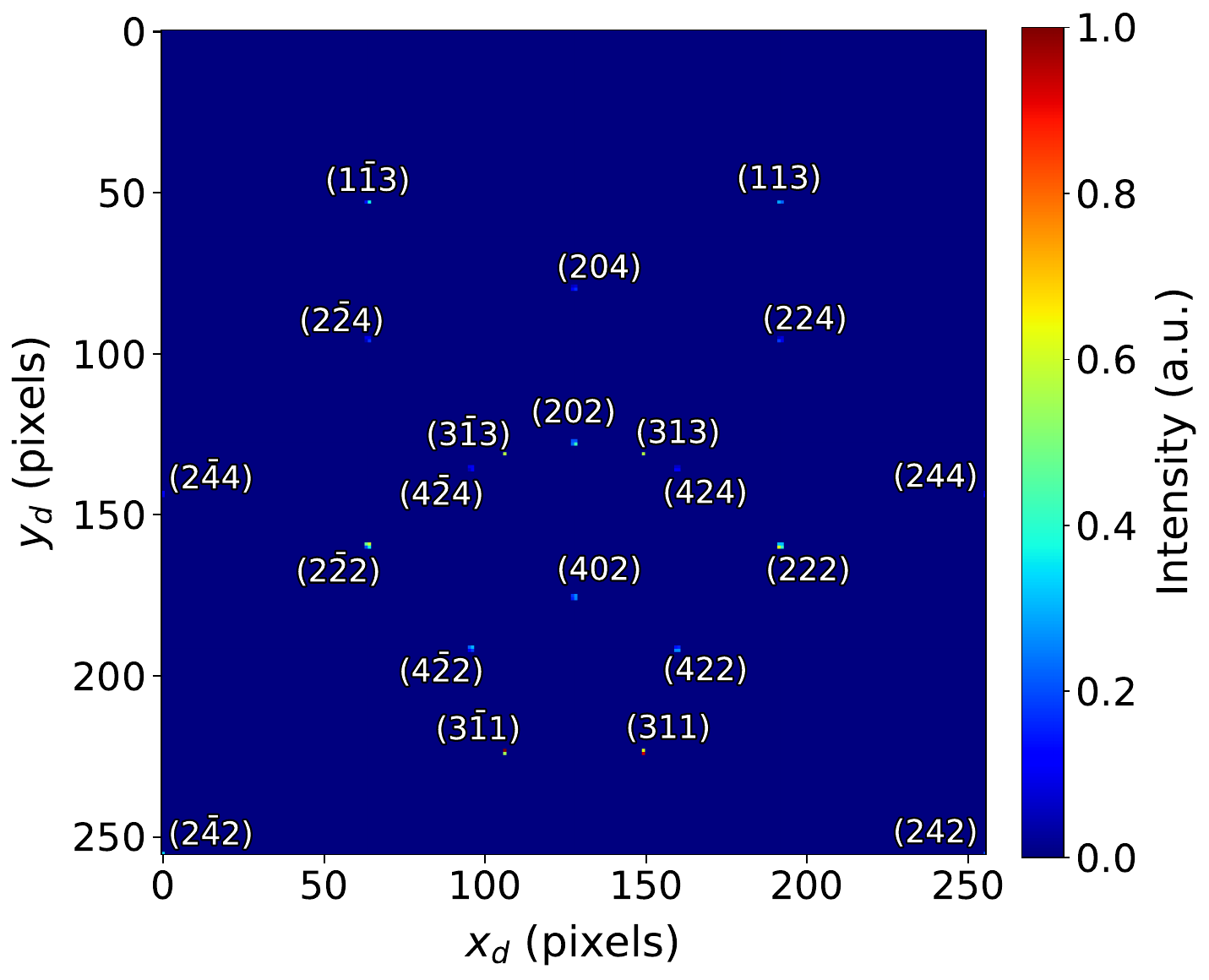}
    \caption{Full simulated Laue diffraction pattern.}
    \label{fig:xrd_full}
\end{subfigure}
\hspace{0.1\textwidth}
\begin{subfigure}[t]{0.38\textwidth}
    \centering
    \includegraphics[height=5.2cm]{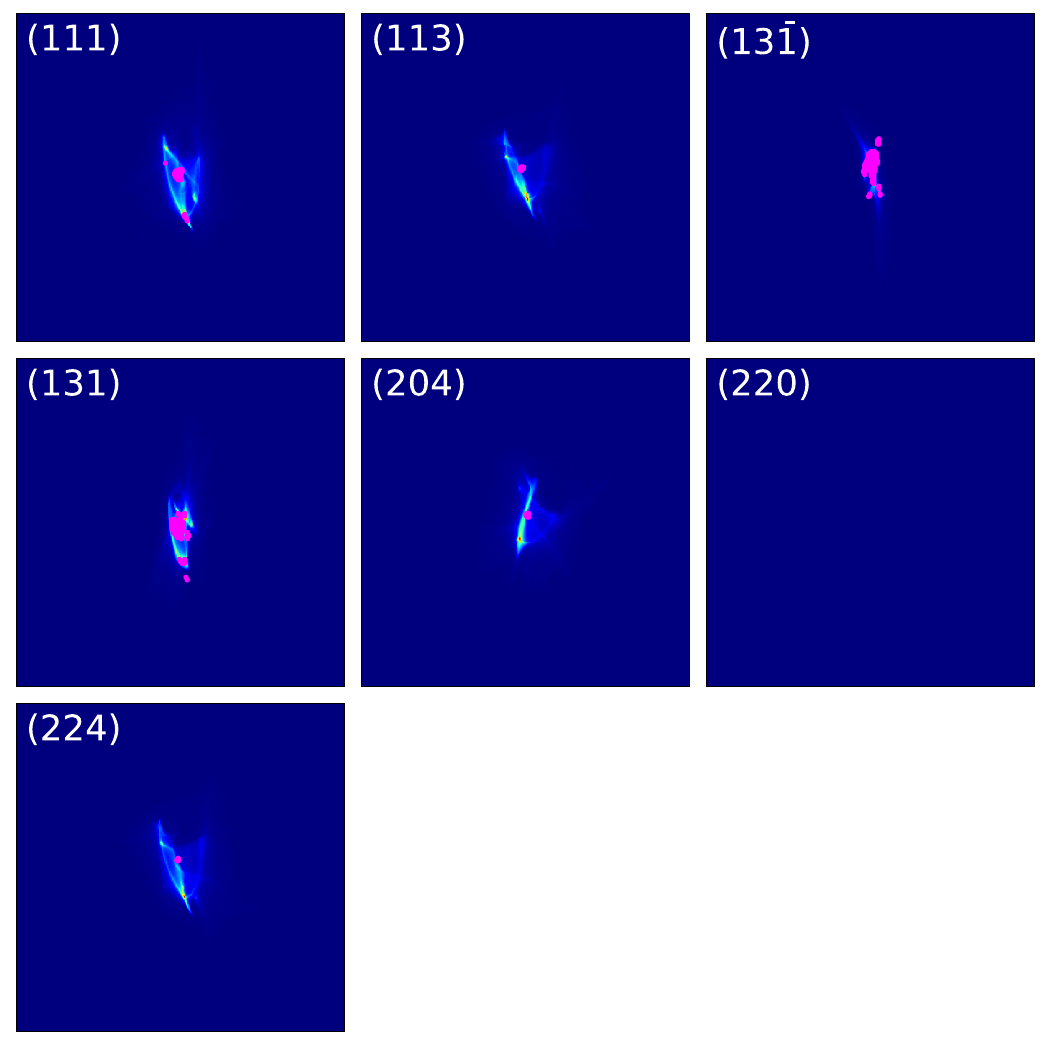}
    \caption{Seven reflections selected for analysis.}
    \label{fig:xrd_peaks}
\end{subfigure}
\caption{
Representative virtual diffraction patterns used for feature extraction. (a) Full simulated Laue diffraction pattern. (b) Selected reflections with retained high-variance pixels highlighted in magenta.
}
\label{fig:xrd_examples}
\end{figure}

The virtual diffraction dataset comprises 10,000 patterns spanning the four loading orientations ($[100]$, $[110]$, $[111]$, and $[234]$) together with the corresponding $\pm\SI{5}{\degree}$ and $\pm\SI{10}{\degree}$ loading-axis misorientations. Each pattern is recorded on a $256 \times 256$ pixel grid for seven selected reflections, resulting in a feature vector of size 458,752 per observation. Representative virtual diffraction patterns are shown in Figure~\ref{fig:xrd_examples}. Figure~\ref{fig:xrd_full} displays the full simulated Laue diffraction pattern generated from a representative DDD dislocation configuration. Figure~\ref{fig:xrd_peaks} shows the seven reflections selected for detailed analysis: [111], [113], [13$\bar{1}$], [131], [204], [220], and [224]. For each reflection, the central region containing the high-variance pixels retained after feature selection is highlighted in magenta, illustrating the diffraction features that exhibit the strongest variability across the full dataset.

\begin{figure}[htbp]
    \centering
    \includegraphics[width=\textwidth]{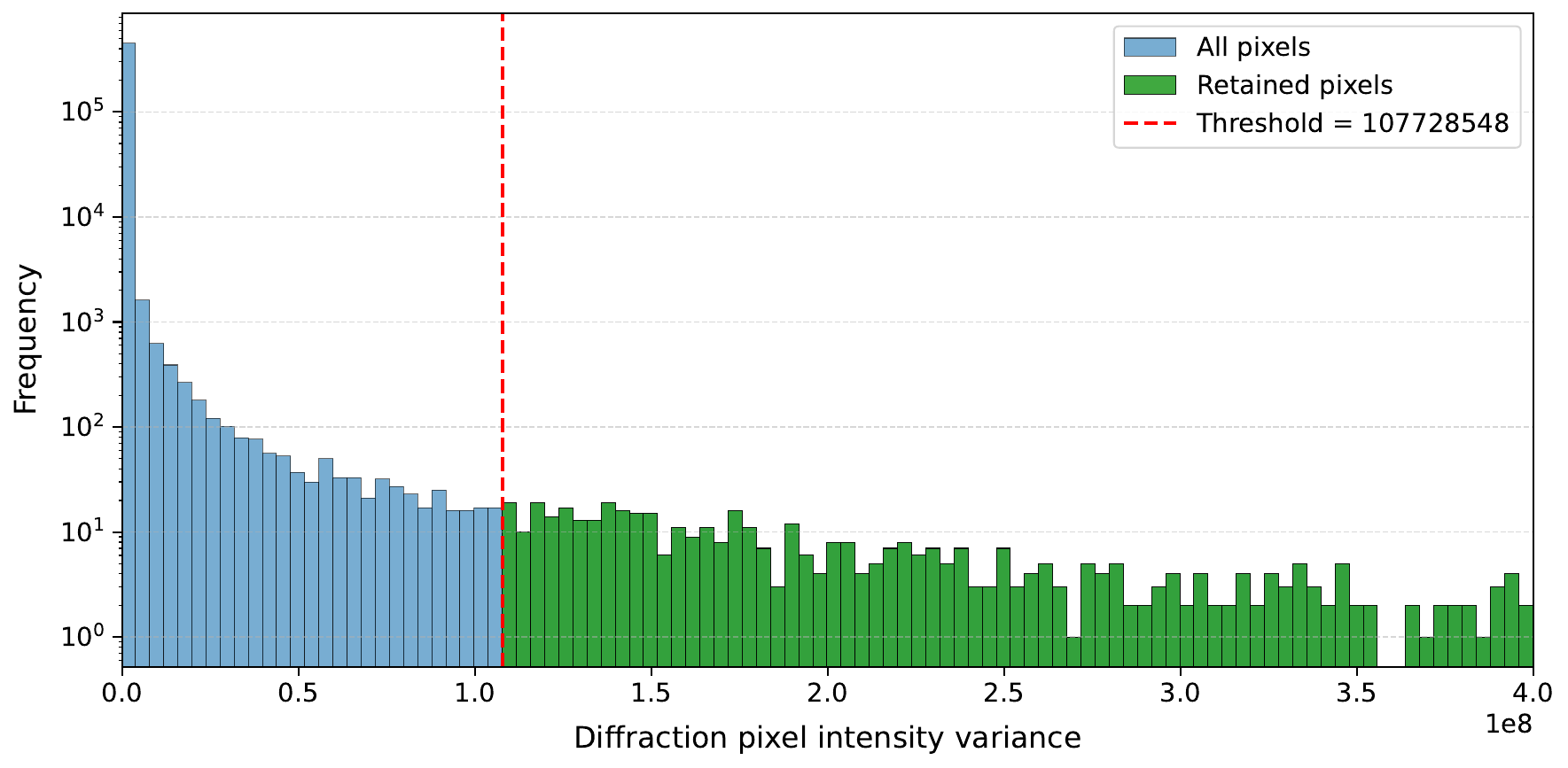}
    \caption{
    Variance-based feature selection applied to the XRD dataset. The histogram shows the variance distribution across all diffraction pixels, with the retained high-variance pixels overlaid in green and the selection threshold indicated by the dashed line. The frequency axis is shown on a logarithmic scale. For visualization clarity, the horizontal axis is limited to the range where the majority of pixel variances are concentrated.
    }
    \label{fig:variance_threshold}
\end{figure}

To reduce the high dimensionality of the diffraction representation, feature selection is applied \cite{guyon2003introduction}. Specifically, the variance of each diffraction pixel is computed across all observations in the dataset. Only pixels with variances exceeding a threshold of $1.08\times10^8$, corresponding to the 99.84th percentile of the variance distribution, are retained, reducing the diffraction representation from 458,752 to 735 features. Figure~\ref{fig:variance_threshold} shows the variance distribution of all diffraction pixels, with the retained high-variance pixels overlaid and the applied selection threshold indicated.

The retained diffraction features correspond to pixels exhibiting the highest variance across the dataset and are paired with the corresponding dislocation density fields to form the multimodal dataset used for cross-modal contrastive learning.

\subsection{Cross-Modal Representation Learning}
\begin{figure}[htbp]
\centering

\begin{subfigure}[t]{0.49\textwidth}
    \centering
    \includegraphics[width=\textwidth]{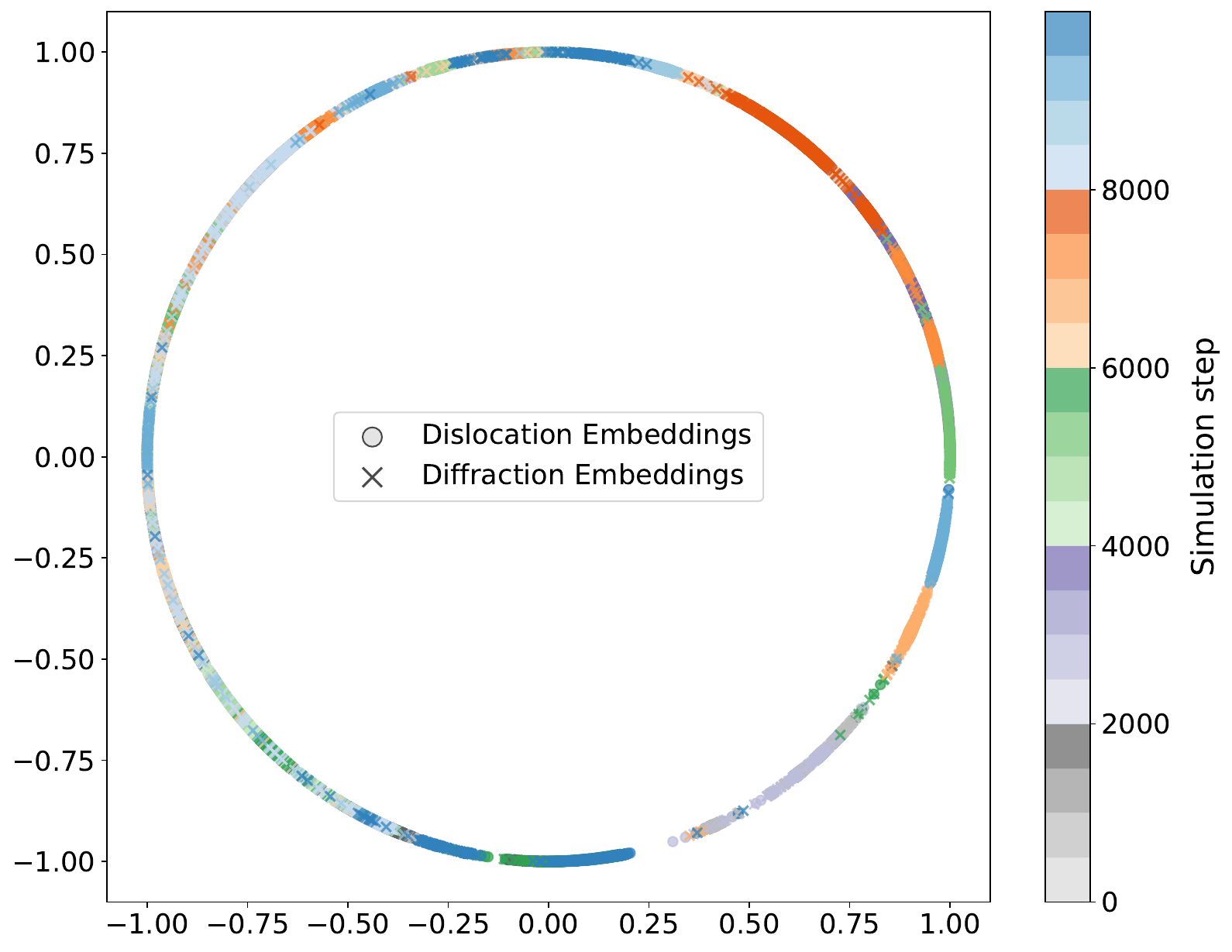}
    \caption{Cross-modal latent embeddings}
    \label{fig:latent_space_embeddings}
\end{subfigure}
\hfill
\begin{subfigure}[t]{0.49\textwidth}
    \centering
    \includegraphics[width=\textwidth]{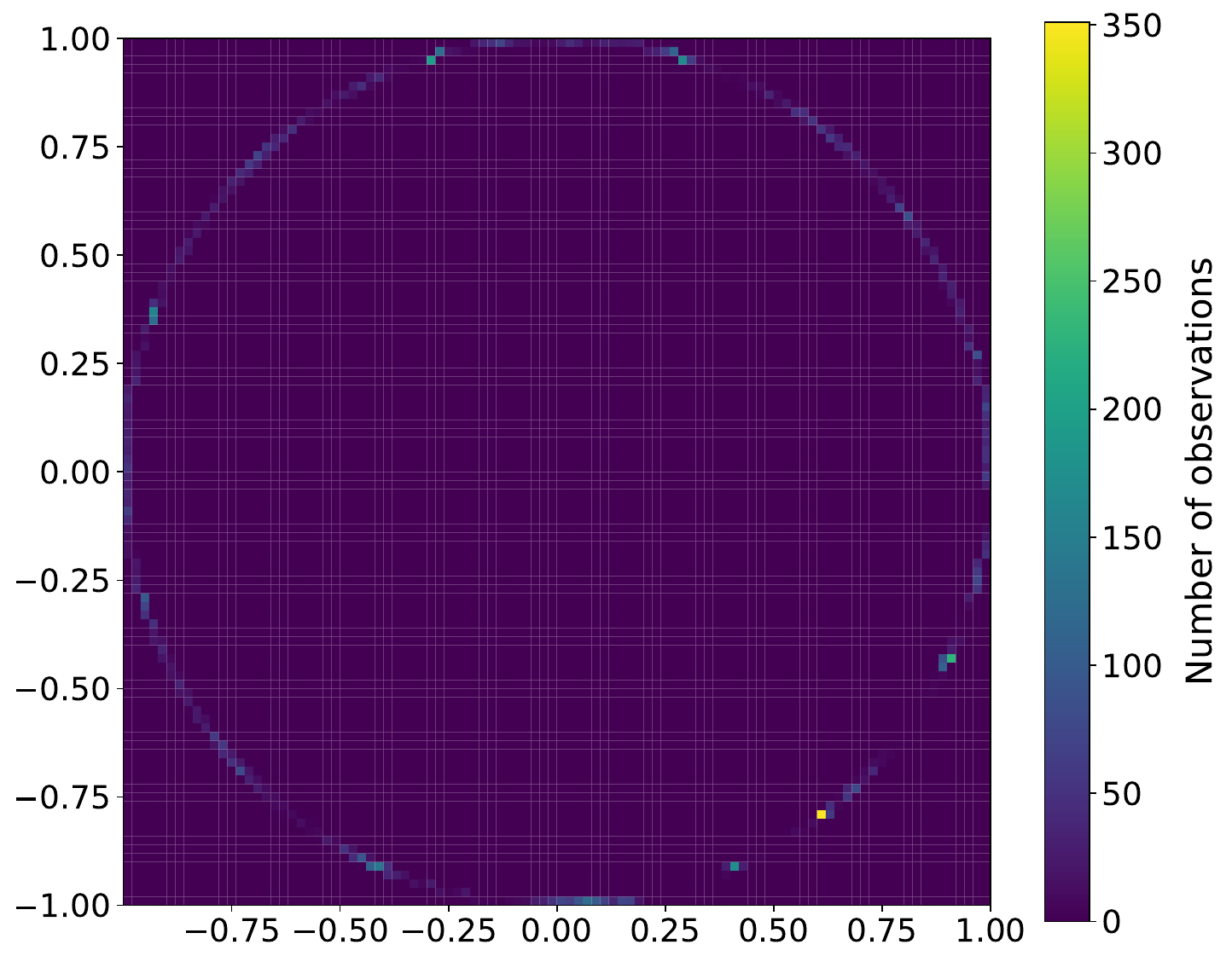}
    \caption{Density of observations}
    \label{fig:latent_space_density}
\end{subfigure}

\caption{
2D latent space learned by the cross-modal contrastive model. (a) Density-field embeddings (circles) and diffraction embeddings (crosses) for all 10,000 observations, with the color scale representing the simulation step. (b) Density map of the density-field embeddings showing the number of observations within each region of the latent space. The density map highlights variations in the distribution of observations that are obscured by overlapping points in (a).
}
\label{fig:latent_space}

\end{figure}

To evaluate the ability of the contrastive learning model to capture structural information across modalities, the latent embeddings obtained from the structural and diffraction representations are analyzed directly in the shared 2D latent space. Figure~\ref{fig:latent_space_embeddings} shows the density-field and diffraction embeddings for all 10,000 paired observations. The density-field embeddings and their corresponding diffraction embeddings occupy closely aligned regions of the latent space, indicating successful cross-modal alignment. As substantial overlap between observations occurs along a circle in latent representation, the distribution of the density-field embeddings is further visualized using a 2D density map in Figure~\ref{fig:latent_space_density}. The density map shows the number of observations within each region of the latent space, revealing variations in observation density that are obscured by overlapping points in the scatter representation in Figure~\ref{fig:latent_space_embeddings}.

\begin{figure}[htbp]
\centering

\begin{subfigure}[t]{0.48\textwidth}
    \centering
    \includegraphics[width=\textwidth]{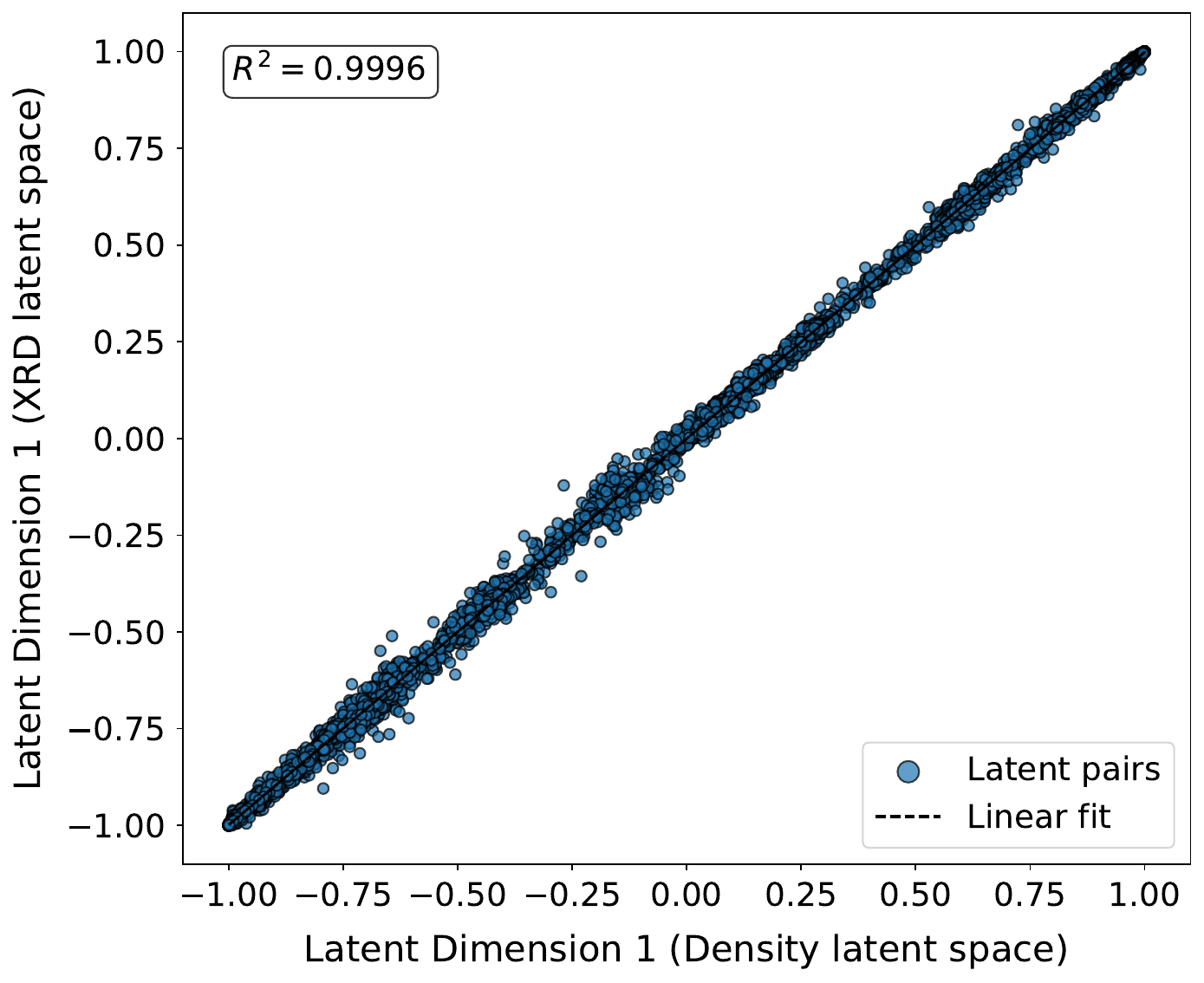}
    \caption{Latent dimension 1 comparison}
    \label{fig:latent_dim1}
\end{subfigure}
\hfill
\begin{subfigure}[t]{0.48\textwidth}
    \centering
    \includegraphics[width=\textwidth]{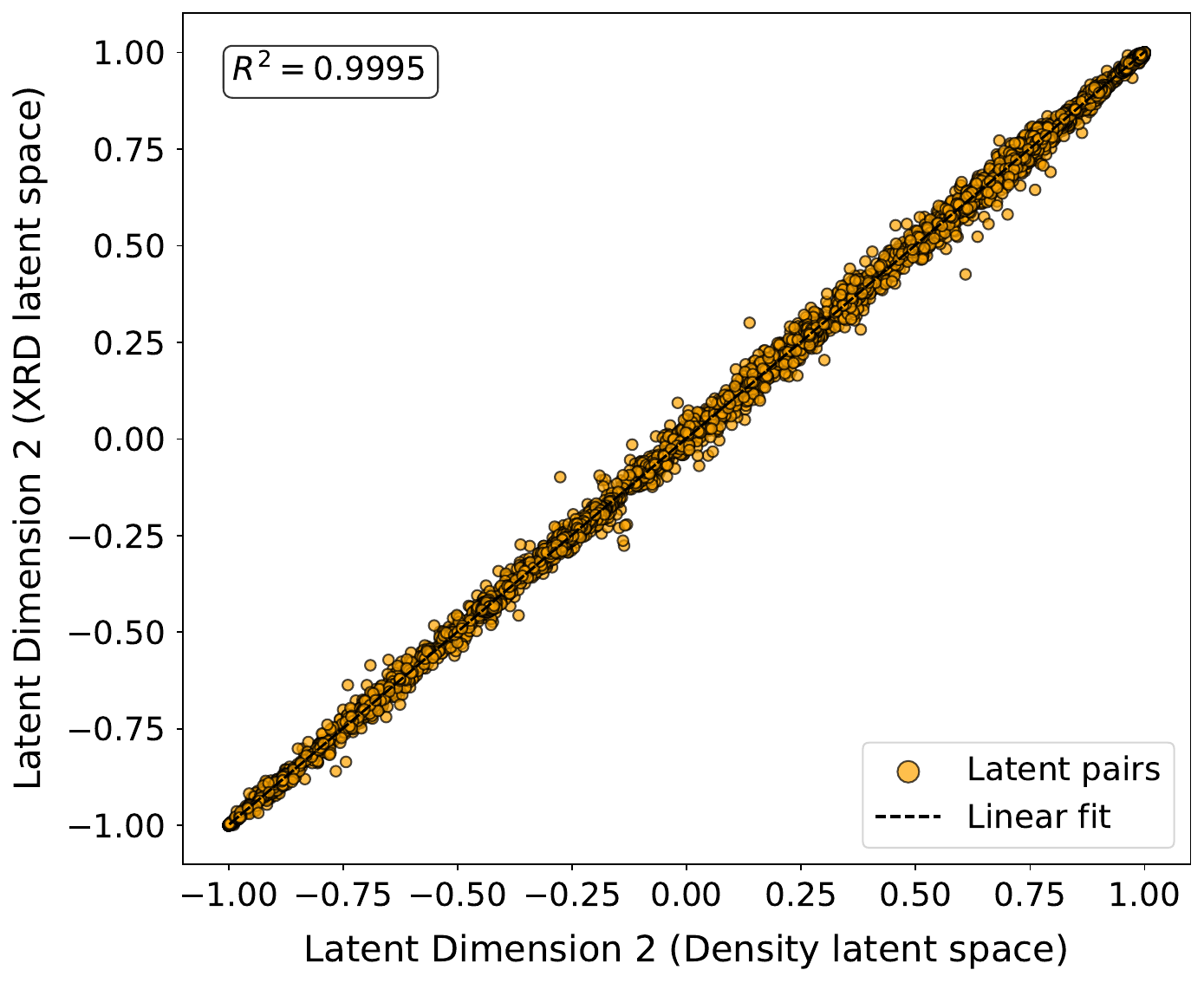}
    \caption{Latent dimension 2 comparison}
    \label{fig:latent_dim2}
\end{subfigure}

\caption{
Comparison of corresponding density-field and diffraction embeddings in the learned 2D latent space. 
(a) Latent dimension 1 and 
(b) latent dimension 2. Each point corresponds to a dislocation microstructure observation represented by its paired density-field and diffraction latent coordinates, with the dashed line indicating the linear regression fit. The near-perfect alignment ($R^2 > 0.999$) highlights the strong linear correspondence between the two modalities in the learned latent space.
}
\label{fig:r2_score}
\end{figure}

This alignment is quantified using the coefficient of determination ($R^2$) between corresponding dimensions of the density-field and diffraction embeddings. Figure~\ref{fig:r2_score} shows the relationship between the two modalities along the two dimensions of the learned latent space. Strong linear correspondence is observed for both latent dimensions, with $R^2=0.9996$ and $R^2=0.9995$ for latent dimensions 1 and 2, respectively.

To determine the minimum amount of structurally diverse data required to effectively train the cross-modal model, the contrastive learning model is trained using subsets of observations selected through farthest point sampling (FPS) in the structural embedding space. FPS is first applied to the full dataset of 10,000 observations to obtain an ordered list of samples that progressively span the structural variability of the dataset. From this ordering, subsets consisting of 10, 50, 100, 500, 1000, 2000, 3000, 5000, 7000, and 9000 observations are considered. For each FPS subset, the observations are partitioned into training and test sets, with 80\% used for training and 20\% reserved for testing.

Model generalization is subsequently evaluated using an additional validation set consisting of the next-1000 observations in the FPS ordering that are not included in the training or test subsets. This strategy ensures that the validation samples correspond to previously unseen regions of the structural manifold.
Because FPS orders observations according to structural diversity, selecting the next-1000 FPS-selected observations constitutes an intentional bias toward a `worst-case' evaluation, as these samples are the most structurally distinct from those used for training. The resulting validation therefore assesses the model’s ability to generalize to maximally distinct structures.

We ultimately quantify model performance by comparing the retrieved density fields with their corresponding ground-truth density fields. For each diffraction pattern in the validation set, the diffraction representation is mapped into the shared latent space, and the nearest structural representation among the available training observations is identified based on cosine similarity. The corresponding density field is then retrieved from the training dataset, together with its associated 3D dislocation microstructure. Thus, retrieval refers to identifying an existing structure from the dataset rather than reconstructing a new density field or dislocation configuration. Because each retrieved density field is associated with an original DDD configuration, the corresponding 3D dislocation structure, including its Burgers-vector information, can be directly looked up from the simulation dataset.

\begin{figure}[htbp]
\centering

\begin{subfigure}[t]{0.496\textwidth}
    \centering
    \includegraphics[width=\textwidth]{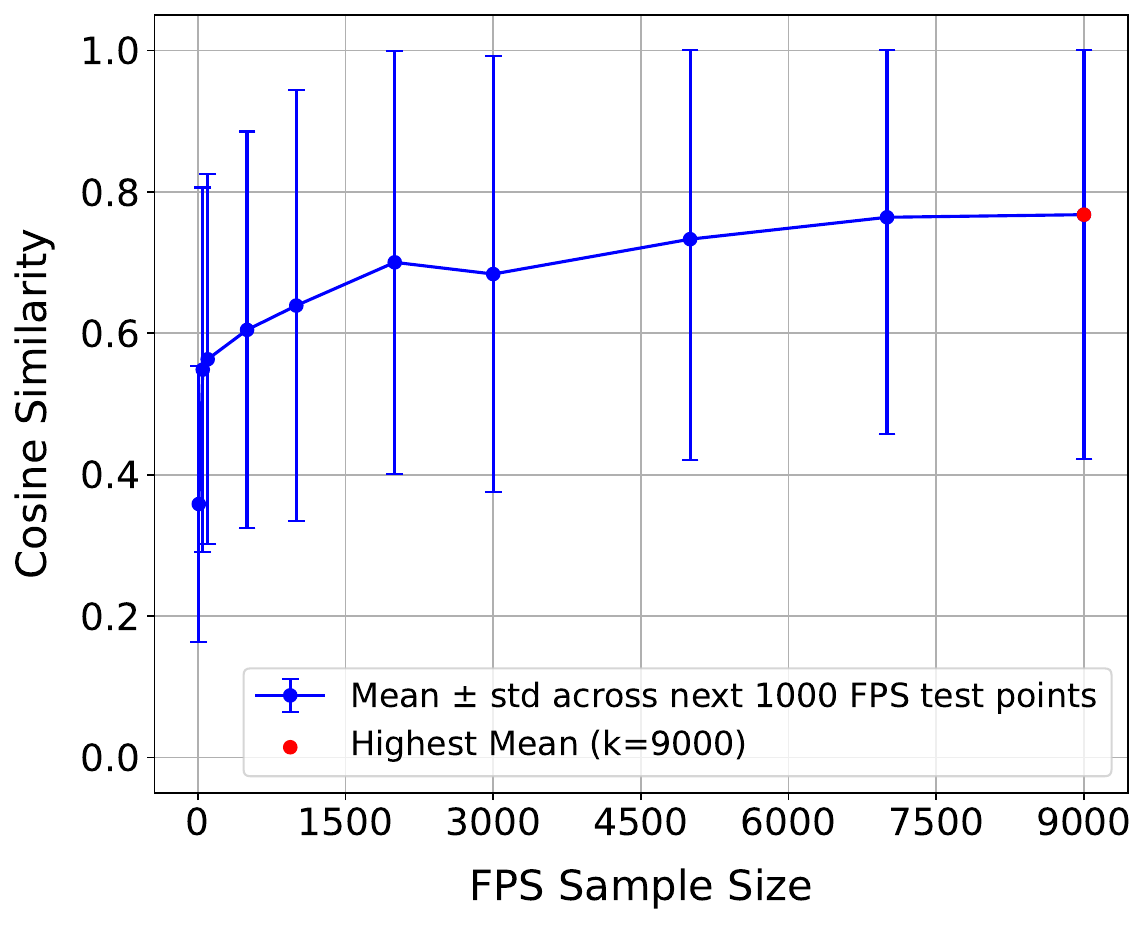}
    \caption{XRD-to-density retrieval performance in the original density-field space.}    
    \label{fig:cs_fps_org}
\end{subfigure}
\hfill
\begin{subfigure}[t]{0.496\textwidth}
    \centering
    \includegraphics[width=\textwidth]{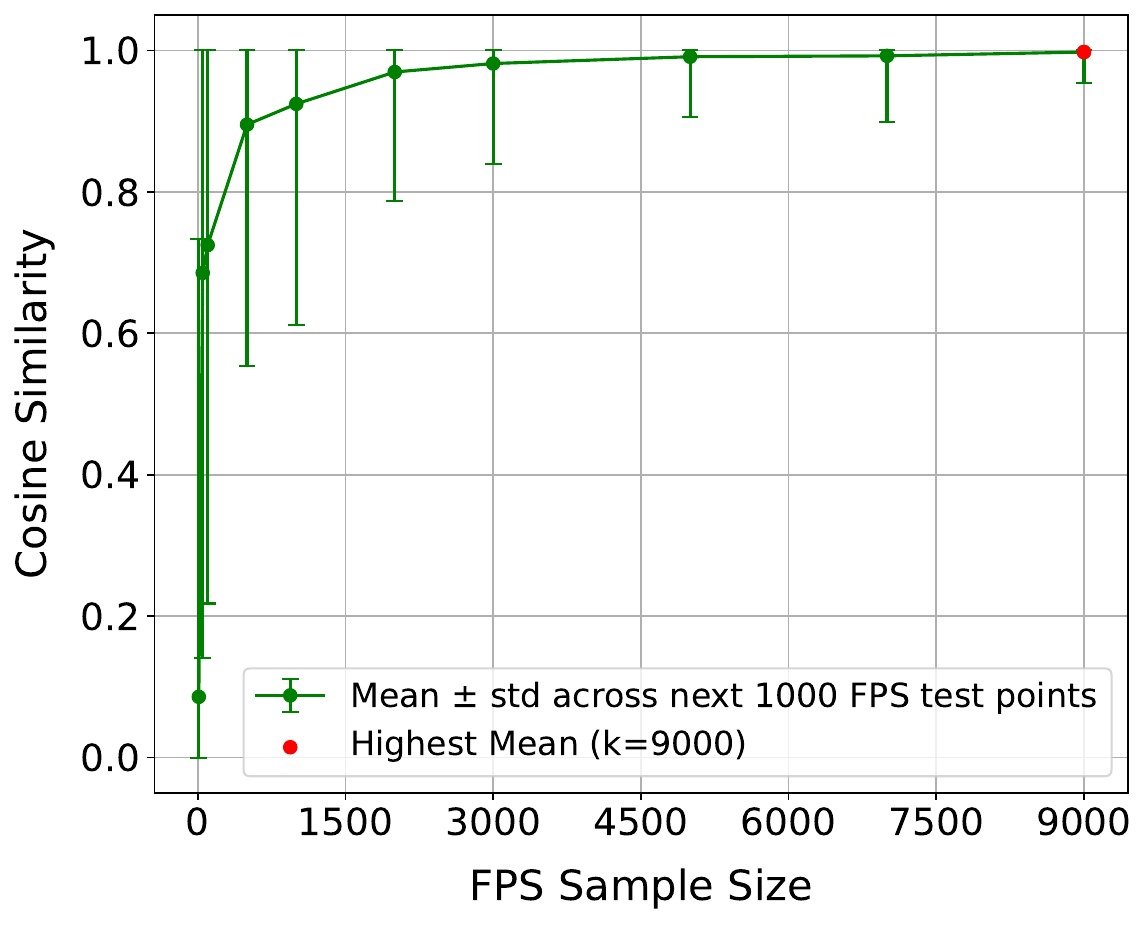}
    \caption{XRD-to-density retrieval performance in the learned latent space.}
    \label{fig:cs_fps_emb}
\end{subfigure}

\caption{
Model performance as a function of FPS training subset size.
(a) Cosine similarity between retrieved and ground-truth structures evaluated in the original density-field space. 
(b) Cosine similarity between retrieved and ground-truth structural representations evaluated in the learned latent structural space. In both cases, performance is evaluated on the next 1000 validation observations in the FPS ordering.
}
\label{fig:cosine_vs_fps}
\end{figure}

The comparison is performed both in the original density-field space (Figure~\ref{fig:cs_fps_org}) and in the learned latent structural space (Figure~\ref{fig:cs_fps_emb}). Cosine similarity between the predicted and corresponding ground-truth representations is computed in each evaluation space for the validation set. For each FPS subset size, the cosine similarity scores across the validation set are averaged to obtain an overall performance metric. As shown in Figure~\ref{fig:cosine_vs_fps}, the mean cosine similarity increases with the size of the FPS-selected subsets in both evaluation spaces. The standard deviation across the validation observations generally decreases with increasing subset size. In the shared latent space (Figure~\ref{fig:cs_fps_emb}), the cosine similarity rapidly approaches unity and reaches a plateau at approximately 500 FPS-selected observations, whereas lower similarity values are obtained in the original density-field space (Figure~\ref{fig:cs_fps_org}).

\begin{figure}[htbp]
\centering


\begin{subfigure}[t]{0.19\textwidth}
    \centering
    \makebox[\linewidth][c]{%
        \begin{tabular}{c}
            \textbf{10 FPS}\\[-1pt]
            \phantom{\textbf{(high similarity)}}            
        \end{tabular}%
    }\\[2pt]
    \includegraphics[width=\textwidth]{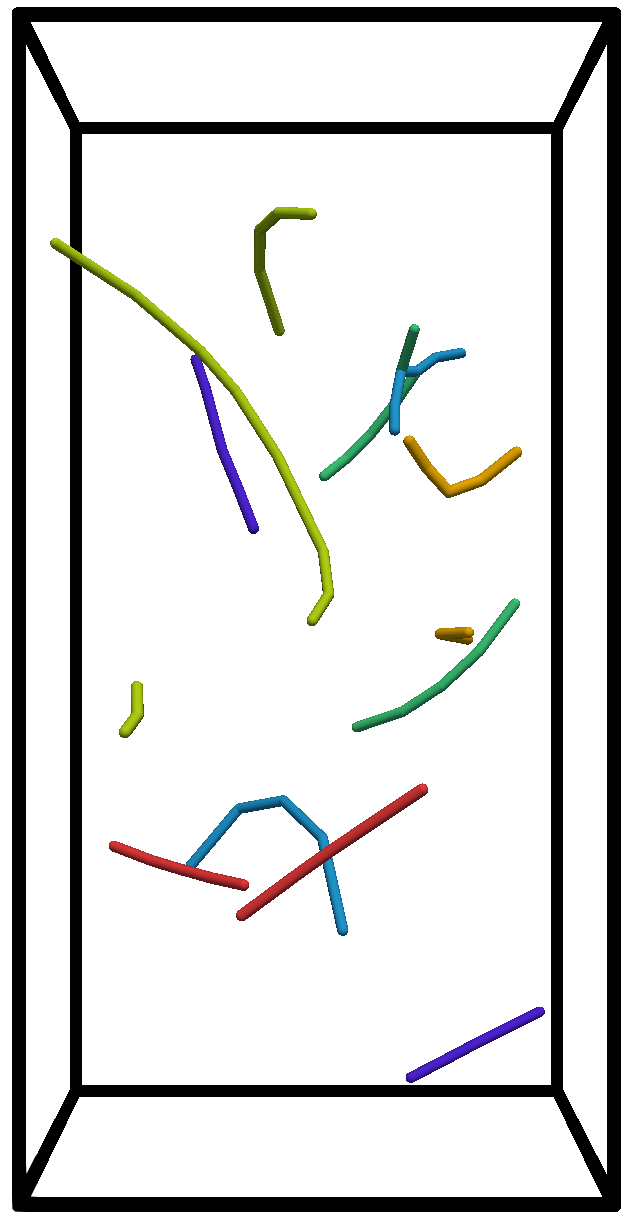}
    \caption{Ground truth}
    \label{fig:FPS_GT_10}
\end{subfigure}
\hspace{0.15\textwidth}
\begin{subfigure}[t]{0.19\textwidth}
    \centering
    \makebox[\linewidth][c]{%
        \begin{tabular}{c}
            \textbf{500 FPS}\\[-1pt]
            \textbf{(high similarity)}
        \end{tabular}%
    }\\[2pt]
    \includegraphics[width=\textwidth]{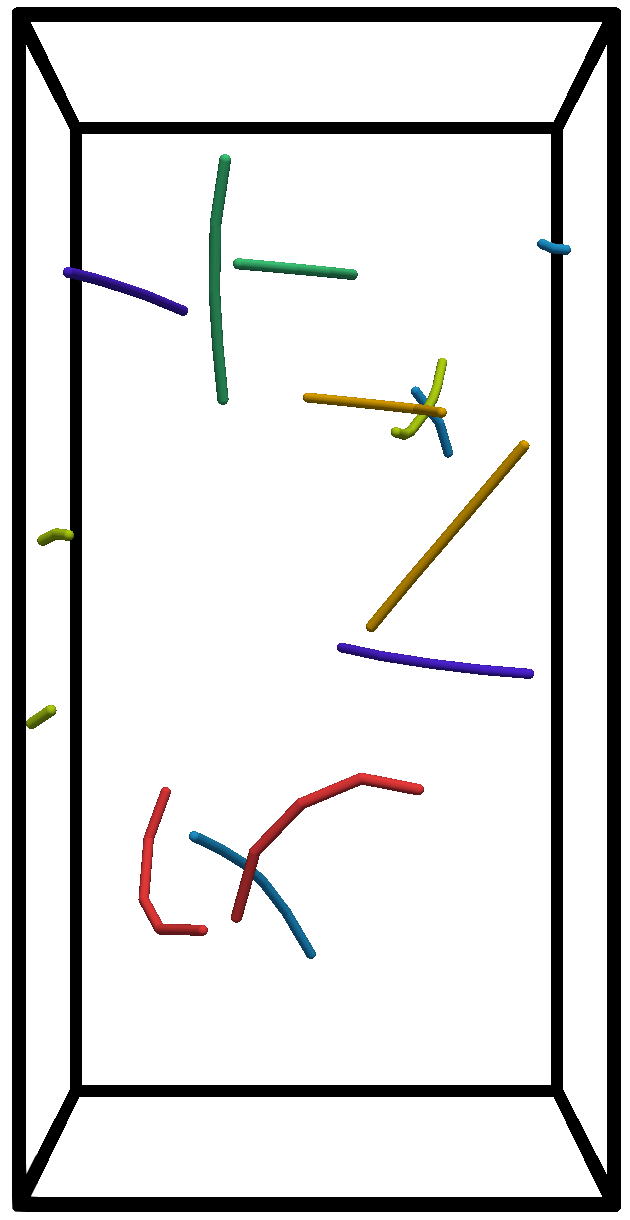}
    \caption{Ground truth}
    \label{fig:FPS_HGT_500}
\end{subfigure}
\hspace{0.15\textwidth}
\begin{subfigure}[t]{0.19\textwidth}
    \centering
    \makebox[\linewidth][c]{%
        \begin{tabular}{c}
            \textbf{500 FPS}\\[-1pt]
            \textbf{(low similarity)}
        \end{tabular}%
    }\\[2pt]
    \includegraphics[width=\textwidth]{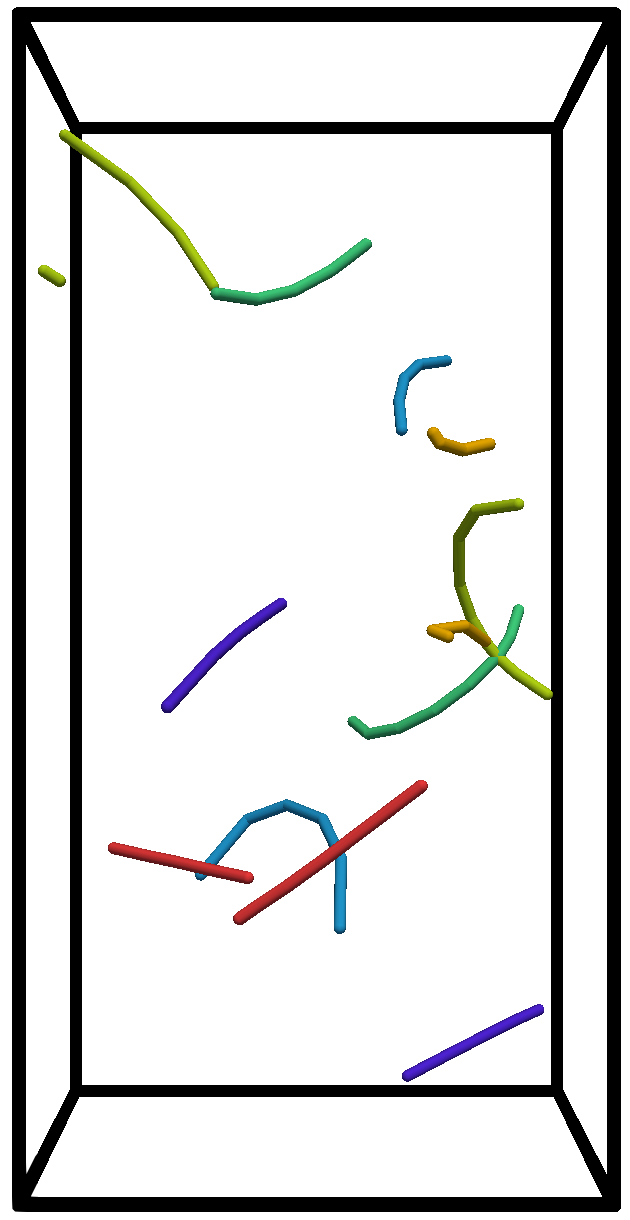}
    \caption{Ground truth}
    \label{fig:FPS_LGT_500}
\end{subfigure}

\vspace{0.3cm}


\begin{subfigure}[t]{0.19\textwidth}
    \centering
    \includegraphics[width=\textwidth]{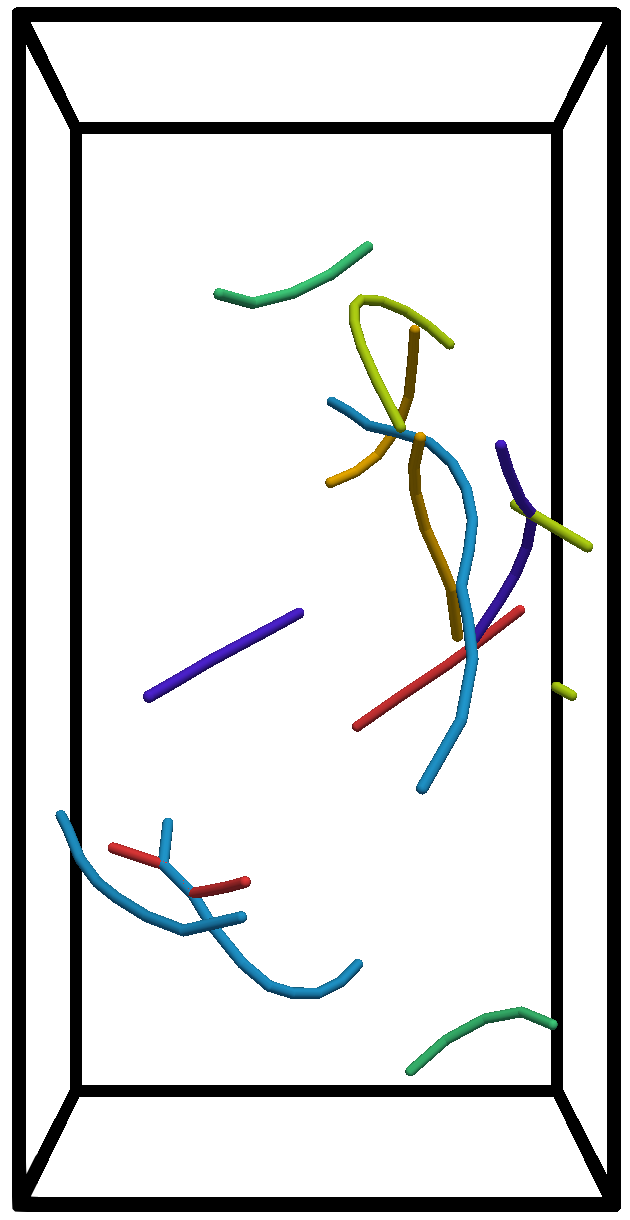}
    \caption{Retrieved}
    \label{fig:FPS_PD_10}
\end{subfigure}
\hspace{0.15\textwidth}
\begin{subfigure}[t]{0.19\textwidth}
    \centering
    \includegraphics[width=\textwidth]{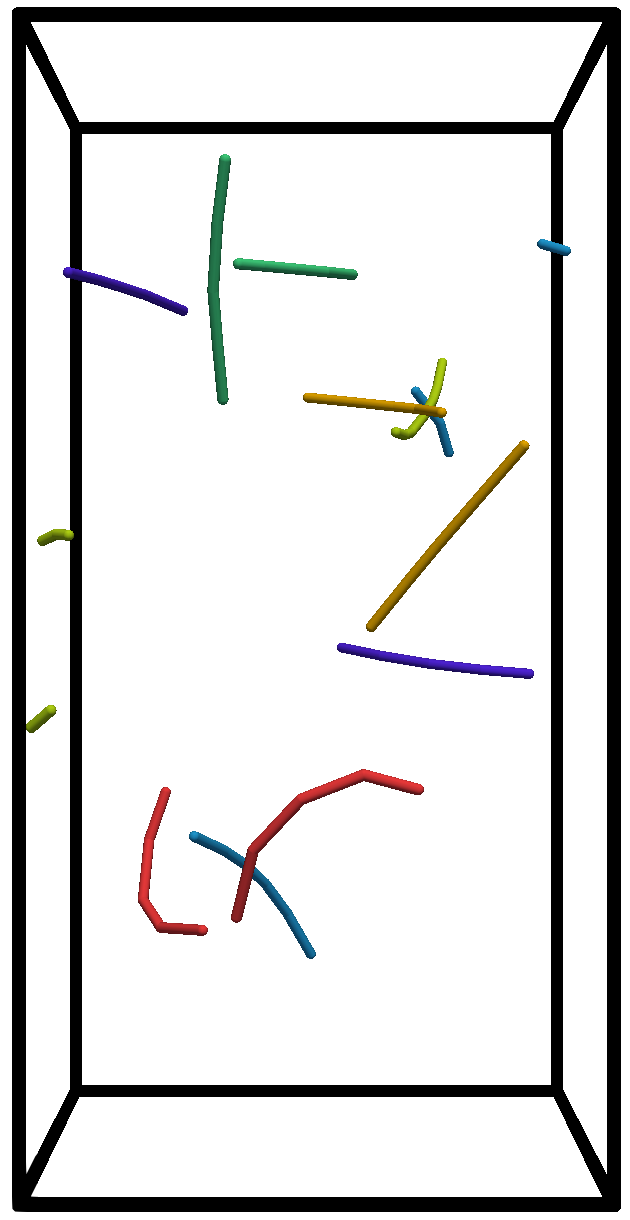}
    \caption{Retrieved}
    \label{fig:FPS_HPD_500}
\end{subfigure}
\hspace{0.15\textwidth}
\begin{subfigure}[t]{0.19\textwidth}
    \centering
    \includegraphics[width=\textwidth]{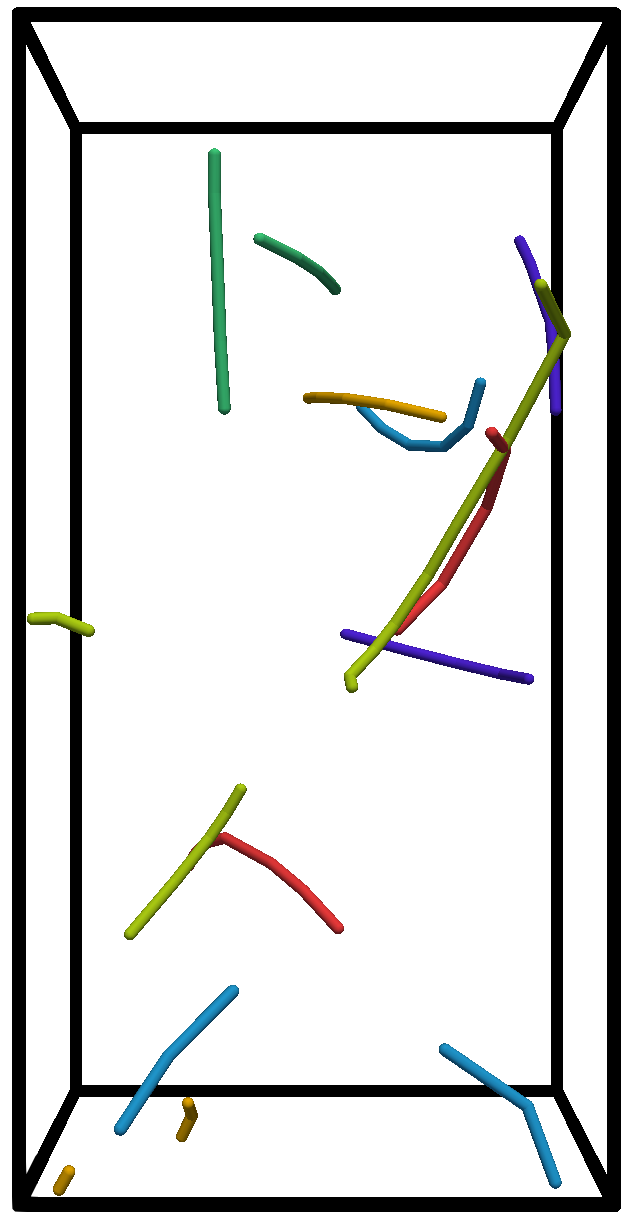}
    \caption{Retrieved}
    \label{fig:FPS_LPD_500}
\end{subfigure}

\caption{
Qualitative comparison between ground-truth (top row) and retrieved (bottom row) 3D dislocation microstructures for representative validation observations. Corresponding ground-truth and retrieved microstructures are arranged vertically. Panels (a,d) show a representative observation from the 10 FPS subset, while panels (b,e) and (c,f) show the observations with the highest and lowest cosine similarity, respectively, among the 1000 validation observations for the 500 FPS subset.
}
\label{fig:structure_prediction}
\end{figure}

Finally, we qualitatively assess the model by visually comparing the 3D dislocation microstructures retrieved through the shared latent space with the corresponding ground-truth microstructures across all FPS subsets. Figure~\ref{fig:structure_prediction} shows representative observations from the 10 and 500 FPS subsets, including the observations with the highest and lowest cosine similarity for the 500 FPS subset. Qualitatively, the retrieved microstructures reproduce the overall spatial distribution of the corresponding ground-truth dislocation configurations, although the agreement in the positions and geometries of individual dislocation segments varies between observations. The strongest agreement is observed for the highest-similarity case within the 500 FPS subset, where the dislocation configurations are nearly identical, while the other cases exhibit greater local differences but retain broadly similar spatial distributions of dislocations within the simulation volume. 

\section{Discussion}\label{sec4}
Our results demonstrate that dislocation structures and their corresponding diffraction signatures can be successfully mapped into a shared latent representation using cross-modal contrastive learning. The close alignment of the structural and diffraction embeddings observed in the shared 2D latent space (Figure~\ref{fig:latent_space}), together with the high $R^2$ values of 0.9996 and 0.9995 between the corresponding latent dimensions (Figure~\ref{fig:r2_score}), confirms that the dominant structural characteristics governing the simulated diffraction response are consistently preserved in the shared latent space.
Consequently, structurally corresponding dislocation density fields and diffraction patterns can be associated through a common latent representation, thereby enabling retrieval of dislocation microstructures directly from diffraction data within the simulated dataset.
Rather than explicitly solving the non-unique inverse diffraction problem, the proposed framework learns this shared representation directly from paired structural and diffraction observations.

An important finding, illustrated in Figure~\ref{fig:cosine_vs_fps}, is that accurate structure prediction does not require exhaustive sampling of the (simulated) dislocation structures. Instead, selecting structurally diverse observations through farthest point sampling provides sufficient coverage of the structural manifold to learn the dominant structure-diffraction relationship. The rapid convergence of prediction accuracy, together with the performance plateau observed for approximately 500 FPS-selected observations, reveals that the essential structural variability of the present dataset is captured by a relatively small representative and diverse subset. This substantially reduces the amount of simulation data required while maintaining high predictive performance. 
However, one caveat remains.
Even though we treat each individual dislocation structure as a single observation, within each trajectory they are correlated and a high total strain configuration can not be simulated independently.
But our result show that the computationally expensive step of forward-simulating the diffraction representation can be limited to a relatively small number of dislocation structures.

Comparison of retrieval performance in the two evaluation spaces shows that cross-modal retrieval in the shared latent structural space consistently outperforms retrieval in the original density-field space. As shown in Figures~\ref{fig:cs_fps_emb} and \ref{fig:cs_fps_org}, retrieval evaluated in the learned latent structural space achieves consistently higher cosine similarity and lower variability across the validation observations than retrieval evaluated in the original density-field space. These results confirm that the learned embeddings preserve the essential structural characteristics of the dislocation structures while providing a robust representation for cross-modal retrieval from XRD patterns. Model performance improves substantially with increasing FPS subset size, with the average cosine similarity increasing from 0.086 for 10 FPS-selected observations to 0.895 for 500 observations and approaching 0.998 for 9000 observations. The largest performance gains are therefore achieved as increasingly representative structures are incorporated into the training set, whereas additional observations beyond approximately 500 provide progressively smaller improvements. This behavior demonstrates that FPS efficiently captures the dominant structural variability of the simulated dataset while substantially reducing the amount of simulation data required for accurate prediction.

Qualitative comparisons between predicted and ground-truth dislocation microstructures provide further insight into the model's predictive capability. As shown in Figure~\ref{fig:structure_prediction}, predictions generated using the 10 FPS subset (Figure~\ref{fig:FPS_PD_10}) exhibit clear discrepancies relative to the corresponding ground-truth structures (Figure~\ref{fig:FPS_GT_10}). In contrast, predictions obtained using the 500 FPS subset generally reproduce the corresponding ground-truth dislocation structures with high similarity, although individual observations exhibit varying levels of agreement. This variability is further illustrated by the predictions corresponding to the highest and lowest cosine similarity among the 1000 validation observations for the 500 FPS subset. The highest-similarity prediction (Figure~\ref{fig:FPS_HPD_500}) closely reproduces its corresponding ground-truth configuration (Figure~\ref{fig:FPS_HGT_500}), whereas the lowest-similarity prediction (Figure~\ref{fig:FPS_LPD_500}) exhibits noticeable deviations from its corresponding ground-truth configuration (Figure~\ref{fig:FPS_LGT_500}), highlighting the remaining variability in prediction quality. These observations are consistent with the cosine similarity analysis and confirm that the learned latent representation captures physically meaningful characteristics of the underlying dislocation structures while enabling reasonably accurate predictions for previously unseen diffraction patterns.

The present framework is intentionally restricted to a simulated dataset, where each diffraction pattern has a unique paired dislocation structure.
Extending the approach to larger and more diverse datasets is expected to introduce increasing ambiguity, as different dislocation configurations may generate similar strain fields and consequently similar diffraction signatures, giving rise to a one-to-many mapping for which no unique solution exists. Under these conditions, the model must identify the most probable structural representation rather than a single deterministic solution. An additional source of non-uniqueness arises from the beam incidence direction.
In the present study, all diffraction patterns are generated using a single fixed beam incidence direction for a single dislocation structure.
In a more general setting, however, the same dislocation structure typically gives rise to different diffraction patterns when probed from different beam incidence directions, resulting in a one-to-many relationship between a given microstructure and its corresponding diffraction signatures. Accounting for this directional dependence would therefore require the framework to associate a single structural configuration with multiple possible diffraction observations.
A likely limitation here is our density field-based representation of the dislocation structure.
Other, more network- and connectivity-aware descriptors~\cite{Weger2021,Starkey2022,Katzer2024} of dislocation structures might prove important.

This ambiguity is expected to become increasingly relevant when extending the framework to \textit{arbitrary} experimental diffraction data. While experimental measurements introduce additional sources of variability, they may also contain subtle contrast arising from instrumental and material effects that is not present in the idealized simulated diffraction patterns considered here. Exploiting such information, together with substantially larger collections of paired structure-diffraction observations, will be essential for extending the present framework to general experimental applications.

\section{Conclusion}
We present a cross-modal contrastive learning framework that establishes a shared latent representation between dislocation structures and their corresponding diffraction patterns. By combining discrete dislocation dynamics simulations, virtual diffraction modeling, manifold learning, and contrastive representation learning, our framework successfully learns the relationship between structural and diffraction modalities. The strong agreement between the learned structural and diffraction embeddings demonstrates consistent alignment between the two modalities, as supported by their close correspondence in the shared 2D latent space and the near-unity $R^2$ values obtained between the corresponding latent dimensions.

The results further demonstrate that model performance improves with increasing dataset size, while representative subsets selected through farthest point sampling efficiently capture the dominant structural variability present in the simulated dataset. In particular, a representative subset of approximately 500 (5\,\%) observations is sufficient to enable accurate retrieval of dislocation microstructures from previously unseen diffraction patterns, reducing the amount of data required for training by approximately 95\% relative to the full dataset of 10,000 observations.

Although the present framework is developed and evaluated using a simulated dataset with uniquely paired dislocation microstructures and diffraction patterns, it provides a foundation for future developments to larger simulated datasets and experimental diffraction measurements. As the diversity of accessible dislocation configurations increases, addressing the inherent ambiguity of the inverse structure-diffraction relationship will become increasingly important. Nevertheless, the proposed framework demonstrates the potential of multimodal representation learning for data-driven interpretation of diffraction measurements and characterization of complex dislocation microstructures.

\section*{Declarations}

\begin{itemize}
\item Funding\\
    B.U. and M.S. gratefully acknowledge funding by the German Research Foundation (DFG) through STR 1729/1-1, Project No. 469020538. MS gratefully acknowledges support by the European Union by ERC grant DISCO-DATA, Project No. 101161287. The views and opinions expressed are, however, those of the authors only and do not necessarily reflect those of the European Union or the European Research Council Executive Agency. Neither the European Union nor the granting authority can be held responsible for them.
    N.B. acknowledges support by the LDRD program (26-ERD-004) at LLNL.
    N.B.'s work was performed under the auspices of the U.S. Department of Energy by LLNL under Contract DE-AC52-07NA27344.
\item Conflict of interest/Competing interests (check journal-specific guidelines for which heading to use)\\
    Not applicable.
\item Data availability\\
    The density field data is available on Zenodo under \url{https://doi.org/10.5281/zenodo.11354118}, the forward-simulated XRD data: \url{https://doi.org/10.5281/zenodo.22289239}
\item Code availability\\
    The code used in this study is available at \url{https://gitlab.ruhr-uni-bochum.de/icams-mids/ddd_xrd}.
\item Author contribution\\
    \textbf{Benjamin Udofia:} Methodology, Software, Validation, Formal analysis, Investigation, Data Curation, Writing - Original Draft, Writing - Review \& Editing, Visualization.
    \textbf{Nicolas Bertin:} Software, Writing - Review \& Editing.
    \textbf{Markus Stricker:} Conceptualization, Methodology, Formal analysis, Resources, Supervision, Writing - Review \& Editing, Project administration.
\end{itemize}

\noindent

\printbibliography

\end{document}